\documentclass{IEEEtran}

\usepackage{cite}
\usepackage{amsmath,amssymb,amsfonts}
\usepackage{bm}              
\usepackage{graphicx}
\usepackage{textcomp}
\usepackage{siunitx}
\usepackage{array}
\usepackage{booktabs}
\usepackage{tabularx}
\usepackage{stfloats}        
\usepackage{balance}

\usepackage[ruled,vlined]{algorithm2e}

\usepackage[caption=false,font=normalsize,labelfont=sf,textfont=sf]{subfig}

\usepackage{xcolor}
\usepackage{listings}

\usepackage{url}
\usepackage{verbatim}
\usepackage{colortbl}

\usepackage{hyperref}
\hypersetup{
    colorlinks=true,
    linkcolor=black,
    filecolor=black,
    urlcolor=blue,
    citecolor=blue,
    pdftitle={Paper Omitted for Review}
}

\def\BibTeX{{\rm B\kern-.05em{\sc i\kern-.025em b}\kern-.08em%
    T\kern-.1667em\lower.7ex\hbox{E}\kern-.125emX}}

\begin{document}
\title{Automated Angular Received-Power Characterization of Embedded mmWave Transmitters Using Geometry-Calibrated Spatial Sampling}

\author{%
Maaz~Qureshi,~\IEEEmembership{Member,~IEEE}, Mohammad~Omid~Bagheri,~\IEEEmembership{Member,~IEEE}, Abdelrahman~Elbadrawy, 
William~Melek,~\IEEEmembership{Senior~Member,~IEEE}, and~George~Shaker,~\IEEEmembership{Senior~Member,~IEEE}%

\thanks{GitHub: \url{https://github.com/Maaz-qureshi98/RAPTAR-Motion-Planning.git}}%

\thanks{M. Qureshi (M.A.Sc. grad-student, m23qures@uwaterloo.ca) and W. Melek (Supervisor, wmelek@uwaterloo.ca) are with the Department of Mechanical and Mechatronics Engineering. M. O. Bagheri (Postdoc Fellow, omid.bagheri@uwaterloo.ca), A. Elbadrawy (M.A.Sc. grad-student, (kmkhairy@uwaterloo.ca), and G. Shaker (Supervisor, gshaker@uwaterloo.ca) are with the Department of Electrical and Computer Engineering, University of Waterloo, ON N2L 3GL, Ontario, Canada.}%
}


\maketitle
\begin{abstract}
This paper presents an automated measurement methodology for angular received-power characterization of embedded millimeter-wave (mmWave) transmitters using geometry-calibrated spatial sampling. Characterization of integrated mmWave transmitters remains challenging due to limited angular coverage and alignment variability in conventional probe-station techniques, as well as the impracticality of anechoic-chamber testing for platform-mounted active modules. To address these challenges, we introduce RAPTAR, an autonomous measurement system for angular received-power acquisition under realistic installation constraints. A collaborative robot executes geometry-calibrated, collision-aware hemispherical trajectories while carrying a calibrated receive probe, enabling controlled and repeatable spatial positioning around a fixed device under test. A spectrum-analyzer-based receiver chain acquires amplitude-only received power as a function of angle and distance following quasi-static pose stabilization. The proposed framework enables repeatable angular received-power mapping and power-domain comparison against idealized free-space references derived from full-wave simulation. Experimental results for a 60-GHz radar module demonstrate a mean absolute received-power error below 2 dB relative to simulation-derived references and a 36.5 \% reduction in error compared to manual probe-station measurements, attributed primarily to reduced alignment variability and consistent spatial sampling. Unlike classical antenna metrology, the proposed approach does not rely on coherent field acquisition, probe correction, or near-/far-field transformations, and instead targets geometry-consistent power-domain characterization of installed active mmWave modules. The methodology is readily applicable to embedded radar systems integrated into vehicles, aerial platforms, and robotic structures—such as automotive bumpers or unmanned aerial vehicles—without loss of generality, providing a practical and portable solution for angular validation when conventional turntables and anechoic facilities are impractical or infeasible.
\end{abstract}


\section{Introduction}\label{sec:I}
\begin{figure}[!t]
\centerline{\includegraphics[width=0.50\textwidth]{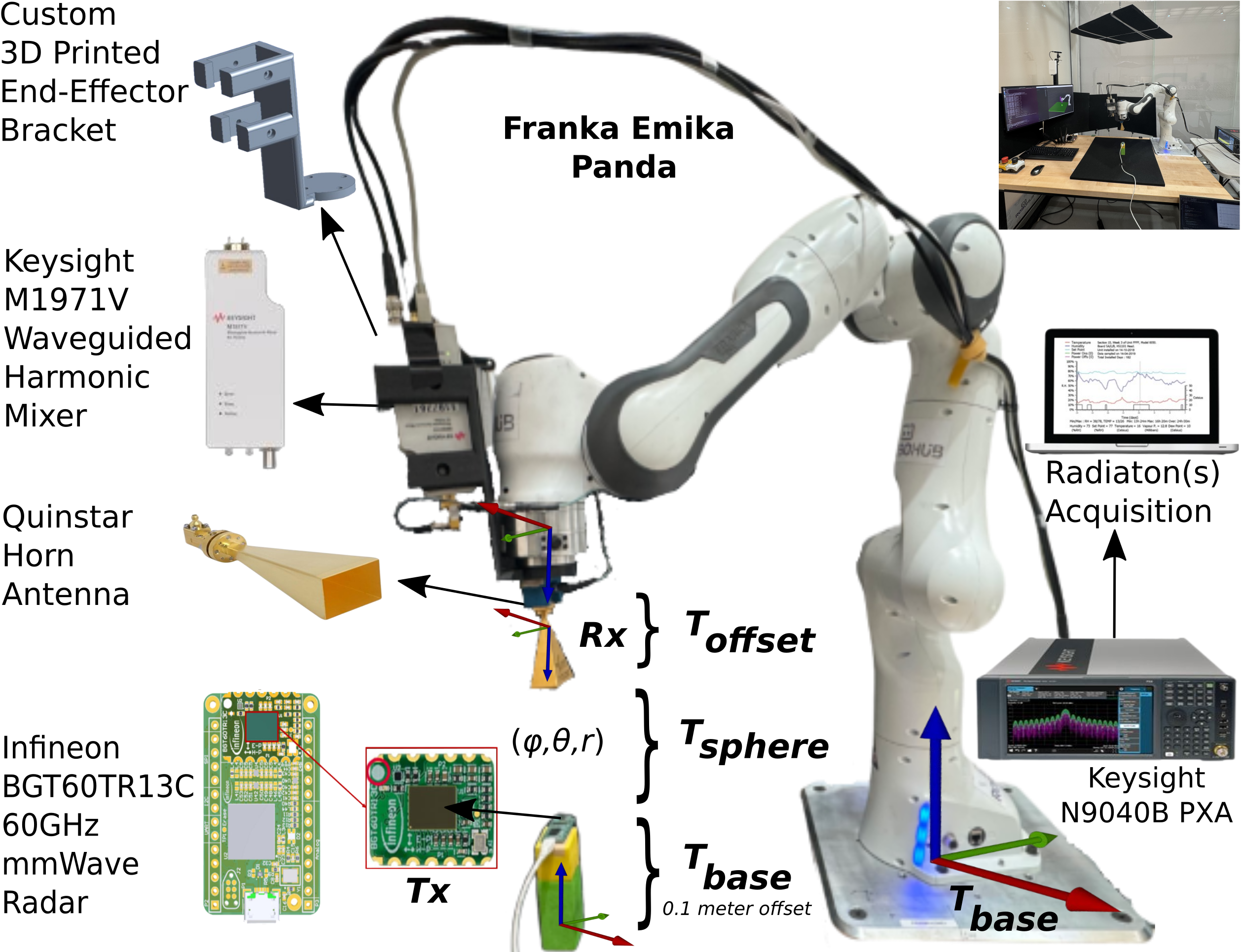}}
\caption{Applied setup overview of the RAPTAR system with attachments.}
\label{fig1}
\end{figure}

\IEEEPARstart{M}{any} practical deployments of embedded millimeter-wave (mmWave) transmitters cannot be characterized using conventional turntables or chamber-based fixtures due to geometric constraints, platform integration, and active electronics, creating a need for alternative angular received-power measurement methodologies. Under such constraints, the measurement problem is no longer one of classical antenna metrology, but rather one of repeatable microwave power measurement with controlled, geometry-calibrated spatial sampling around the device under test. The contribution of this work therefore lies in measurement methodology and repeatable spatial sampling under realistic constraints. In practice, validating the angular behavior of embedded mmWave transmitters at the system level relies on reliable measurement of received power as a function of spatial orientation and separation, rather than on idealized field quantities. For active radar modules integrated with on-chip antennas, packaging, power delivery, and
surrounding structures strongly influence the measured angular response, making
classical antenna metrics difficult to isolate in realistic deployment scenarios. As a
result, practical characterization increasingly focuses on repeatable angular
received-power measurements that capture installation-dependent behavior relevant to
sensing, perception, and communication performance
\cite{richards2005fundamentals,haykin2006cognitive,bagheri2024radar,bagheri2025dielectric,11293594,qureshi2025intelligent}.



As radar systems increasingly transition from discrete RF front ends to tightly
embedded modules, measured performance becomes inseparable from surrounding platform
geometry, materials, and installation constraints. Recent advances in radar-based
sensing have enabled a wide range of emerging applications that rely critically on the
angular response of compact, embedded mmWave transmitters. In automotive and autonomous
systems, mmWave radar is a core sensing modality for perception, localization, and
collision avoidance, where installation-induced angular redistribution of received
power and platform scattering can affect detection performance and robustness
\cite{hasch2012millimeter,waldschmidt2021automotive,patole2017automotive}. Similarly,
radar-based sensing has gained increasing attention in biomedical and wearable
applications \cite{bagheri2024radar,bagheri2024near}, including non-invasive
physiological monitoring \cite{mercuri2013analysis,bagheri2024metasurface}, cardiac
health monitoring \cite{gharamohammadi2025smart}, motion tracking \cite{gu2019motion},
and short-range imaging \cite{brisken2018recent,naghibi2020near}, where device
miniaturization and proximity to lossy media introduce strong coupling and
installation-dependent effects. Across these domains, modules are typically embedded
within complex mechanical structures or operated near lossy materials.

Unlike classical antenna characterization scenarios, embedded radar modules are rarely
operated in ideal free-space conditions \cite{abedi2024use}. Instead, they are commonly
integrated within vehicle bumpers, robotic housings, wearable devices, unmanned aerial
vehicles, and consumer electronics, where dielectric covers, structural supports,
fastening hardware, adhesive layers, cable routing, and nearby scatterers can
substantially alter the angular received-power response. These installation-dependent
effects can distort directional power distributions and introduce platform-specific
scattering mechanisms that are difficult to predict accurately without empirical
validation and often cannot be captured using isolated simulations alone
\cite{hansen2009phased,hasch2012millimeter}. Consequently, there is a growing need for
practical measurement methodologies that can quantify installed angular received-power
behavior under realistic constraints and with repeatable spatial sampling.

Conventional antenna measurement infrastructures---including anechoic chambers, compact
ranges, mechanical turntables, and scanner-based facilities---provide high measurement
fidelity when the device under test (DUT) can be isolated and mounted within a
controlled environment \cite{balanis2015antenna,kraus2002antennas,hansen2009phased}.
However, these approaches impose restrictive fixturing and alignment requirements,
require substantial infrastructure, and assume controlled boundary conditions that are
difficult to maintain for realistic embedded configurations and platform-mounted
active modules \cite{pozar2012microwave,8606248}. Many practical deployments therefore
cannot leverage classical facilities directly, motivating alternative angular measurement methodologies that operate under geometric and installation
constraints.

To address these limitations, recent studies have explored integrating robot-assisted
spatial sampling into RF and mmWave measurement workflows. Robot manipulators enable
repeatable positioning and the execution of complex three-dimensional trajectories in
cluttered environments, making them attractive for constrained sampling around embedded
devices. Early efforts established feasibility primarily through simulation-based or
laboratory-oriented prototypes. \cite{parini2023simulation} proposed a
simulation-based framework for robotic scanning and signal-processing-based retrieval,
but did not address practical deployment issues such as geometry calibration, probe
orientation stability, or real-time robot--instrument coupling. \cite{novotny2017multi}
introduced the LAPS architecture using two coordinated six-DoF manipulators on a
linear rail; while enabling dynamic over-the-air testing, multi-arm synchronization,
calibration overhead, and safety constraints complicate deployment in compact
laboratory and field-adjacent settings. \cite{moser2024rapid} reports robotic
probe alignment using iterative surface fitting and inverse kinematics, but required
dense 3D calibration and external pose tracking systems, reducing portability and
increasing cost.

Commercial industrial measurement platforms further illustrate both the promise of
robotic automation and the associated practical limitations. Robotic antenna
measurement systems developed by NSI MI Technologies \cite{nsi_mi_overview,nsi2025robotic} and Boeing's Dual Robotic Antenna Measurement System (DRAMS) \cite{etslindgren_drams} enable high-precision multi-axis scanning with advanced instrumentation integration. Nevertheless, these platforms are primarily designed for large-scale aerospace hardware and typically require substantial infrastructure, dedicated facilities, and significant capital investment. As a result, their scale and operational complexity render them impractical for rapid prototyping workflows. Collectively, these prototypes illustrate the promise of robotics for
RF testing, yet fall short of providing an integrated, deployable methodology for
mmWave angular pattern characterization of embedded modules under realistic
installation constraints, used in consumer, automotive, robotic, and wearable applications.

In parallel with robotic automation, prior work has also explored semi-controlled
measurement environments that reduce reliance on full anechoic facilities. \cite{4463566} demonstrated that ferrite-based absorber configurations can
emulate controlled fading conditions suitable for wireless device evaluation, while
\cite{8471434} showed that strategic absorber placement can substantially reduce
multipath distortion in non-dedicated measurement spaces. More recently,
\cite{Breinbjerg2023AntennaMC} introduced a semi-anechoic multiprobe architecture
enabling accelerated automotive testing without requiring a full anechoic chamber.
Collectively, these studies confirm that, when properly designed, semi-controlled
environments can yield consistent and reliable power-domain measurement results.
However, these approaches do not incorporate robot-assisted hemispherical spatial
sampling, deterministic motion--measurement coupling, geometry-calibrated end-of-arm
tooling, or fully automated acquisition workflows.

In this paper, we introduce RAPTAR (Radar Radiation Pattern Acquisition through Automated Collaborative Robotics), a fully autonomous measurement system for automated angular received-power characterization of embedded mmWave transmitters under realistic installation constraints. Building on these insights, collaborative robot (cobot) motion planning is employed solely to achieve controlled, repeatable, and geometry-accurate spatial positioning during mmWave measurements, rather than as a sensing or decision-making element. As shown in Fig. \ref{fig1}, The proposed system integrates geometry-calibrated hemispherical spatial sampling, collision-aware motion planning, and time-aligned, pose-indexed RF data acquisition to produce angularly indexed received-power measurements across multiple probe radii within a semi-controlled measurement environment.

\begin{figure*}[t!]
    \centering
    \includegraphics[width=0.99\textwidth]{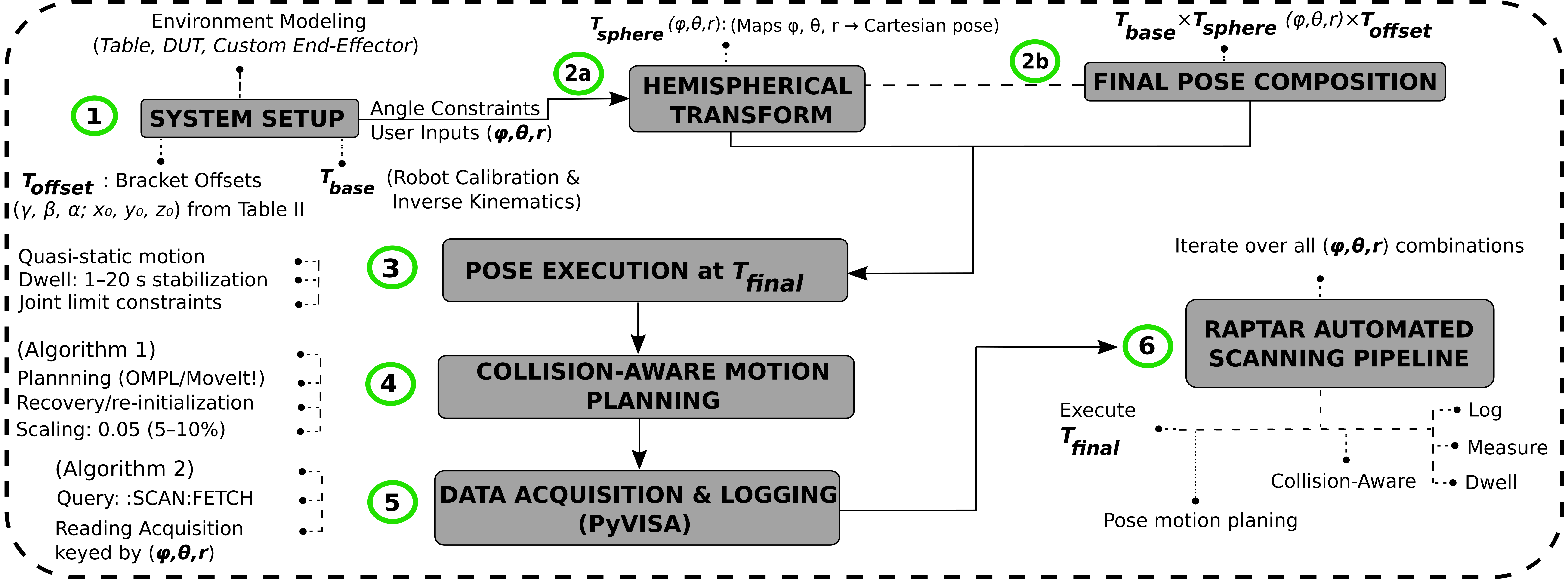}
    \caption{Overview of the RAPTAR measurement pipeline illustrating geometry-calibrated hemispherical spatial sampling, robotic pose execution, and time-aligned acquisition of angularly indexed received-power measurements.}
    \label{fig2}
\end{figure*}

The major contributions of this work are summarized as follows:
(i) A unified microwave measurement methodology for autonomous hemispherical angular
received-power characterization of embedded mmWave modules under realistic
installation conditions, integrating geometry-calibrated spatial sampling with
environment-aware operation and RF data acquisition.
(ii) A deterministic time-aligned acquisition pipeline for pose-indexed received-power
sampling, ensuring repeatable association between executed robot poses and the RF
readout following quasi-static pose stabilization.
(iii) A calibrated, orientation-stable end effector for high-frequency probe control,
with a geometric stability model that maintains probe orientation under multi-axis
trajectories to support repeatable received-power measurements.
(iv) A semi-anechoic, absorber-informed measurement methodology enabling reliable
power-domain measurements outside full anechoic chambers.
(v) Experimental validation using a 60-GHz radar module, demonstrating a mean absolute
received-power error below 2~dB relative to power-domain references derived from
full-wave simulation and up to a 36.5\% reduction in mean absolute error compared to
manual probe-station measurements. Observed reductions in error are attributed
primarily to reduced alignment variability and consistent spatial sampling.

The demonstrated accuracy, repeatability, and operational resilience establish the proposed methodology as a practical approach for engineering validation and comparative characterization of embedded mmWave transmitters under realistic deployment constraints, where conventional chamber-based testing and manual probe positioning are often impractical. In this work, experimental validation focuses on amplitude-only received-power characterization, which is sufficient to quantify installation-dependent angular response distortion and to benchmark against simulation-derived power-domain references and manual baseline measurements. The resulting measurement framework enables rapid, geometry-consistent pattern acquisition with minimal operator intervention, making it well suited for iterative and in situ performance assessment. Extensions to broader measurement modalities require different instrumentation and are left for future work.


\section{System Architecture and Geometric Formulation}
\label{sec:II}

The RAPTAR methodology constitutes a geometry-calibrated mmWave measurement system that leverages robotic motion solely to achieve autonomous, controlled, and repeatable spatial positioning for angular received-power characterization of embedded radar modules. The system is adaptable to a range of embedded platforms by accounting for DUT dimensions, mounting constraints, and operating frequency when selecting the sampling radius, angular resolution, and reachable workspace. Its architecture unifies calibrated spatial sampling with time-aligned RF acquisition to enable repeatable hemispherical measurements without reliance on full anechoic chambers. Synchronization refers to deterministic sequencing of robot pose execution and RF acquisition, not to electromagnetic phase coherence. This section describes the system workflow and the geometric formulation that links commanded sampling coordinates to physical probe poses and pose-indexed received-power samples.

In this work, the framework is experimentally validated using the Infineon BGT60TR13C mmWave radar module, operating over the 58--63~GHz frequency range, as a representative low-profile, high-sensitivity embedded radar platform that has recently seen widespread adoption in automotive and biomedical sensing applications.

\begin{figure}[t!]
\centerline{\includegraphics[width=0.5\textwidth]{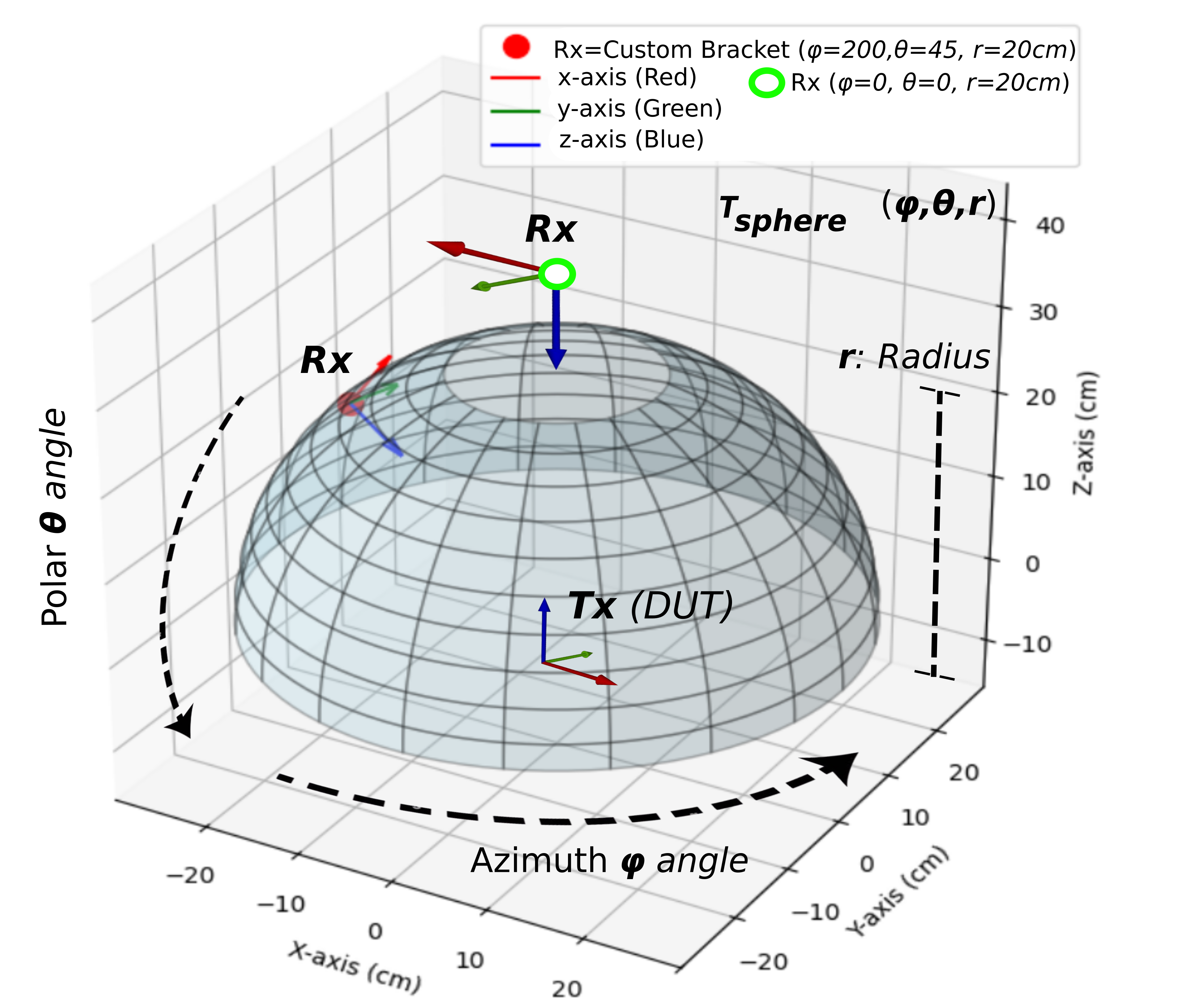}}
   \caption{Three-dimensional sampling grid illustrating the hemispherical measurement geometry in azimuth ($\phi$), elevation ($\theta$), and radius ($r$). The DUT operates as the transmitter, while a receiving antenna is spatially positioned by the collaborative robot to acquire angularly indexed received-power measurements.}  
   \label{fig3}  
\end{figure}

Figure~\ref{fig2} presents the end-to-end workflow, illustrating how user-defined
sampling parameters are mapped to executable probe poses and associated pose-indexed
received-power measurements. The process begins with system setup (Step~1), where the
DUT, supporting structures, and calibrated end-effector are represented in a common
geometric model, and the base transformation $\mathbf{T}_{\mathrm{base}}$ and
end-effector offset $\mathbf{T}_{\mathrm{offset}}$ are established.

In Step~2, spherical sampling coordinates $(\phi,\theta,r)$ are mapped to Cartesian
space through the hemispherical transform $\mathbf{T}_{\mathrm{sphere}}(\phi,\theta,r)$,
defining the sampling direction and radius relative to the DUT. These transforms are
composed to generate the executable end-effector pose,
\begin{equation}
\mathbf{T}_{\mathrm{final}}
=
\mathbf{T}_{\mathrm{base}}
\times
\mathbf{T}_{\mathrm{sphere}}(\phi,\theta,r)
\times
\mathbf{T}_{\mathrm{offset}},
\label{eq:Tfinal}
\end{equation}
which enforces the desired measurement radius $r$ and consistent probe orientation
with respect to the DUT across the hemispherical domain. As a result, each triplet
$(\phi,\theta,r)$ maps to a unique physical probe position on a well-defined sampling
surface, yielding deterministic pose-indexed measurements of received-signal magnitude.
The resulting dataset supports angular received-power mapping, comparison across radii,
and power-domain comparison against idealized free-space references derived from
full-wave simulation.

Step~3 corresponds to quasi-static pose execution, during which the receiving antenna
is positioned at $\mathbf{T}_{\mathrm{final}}$ and held stationary to reduce
measurement variability at millimeter-wave frequencies. In Step~4, collision-aware
motion planning verifies pose reachability within the workspace. Step~5 triggers RF
data acquisition associated with the corresponding $(\phi,\theta,r)$ coordinates.
Finally, Step~6 iterates this sequence over the hemispherical grid to produce a dense
three-dimensional received-power dataset for repeatability assessment and comparative
validation.

Figure~\ref{fig3} illustrates the hemispherical sampling geometry used for angular
received-power measurements. The DUT is fixed at the origin of the measurement
coordinate system, while the receiving antenna is positioned on a hemispherical domain
defined by spherical coordinates $(\phi,\theta,r)$. The azimuth angle $\phi$ specifies
rotation about the vertical axis, the polar angle $\theta$ controls elevation relative
to the DUT boresight, and the radius $r$ defines the measurement distance. For each
sampling point, the receiving antenna is oriented toward the DUT using a consistent
probe reference point to ensure repeatable angular referencing across the sampling
surface.

The angular sampling resolution is selected as a practical trade-off between scan time
and measurement density. Since the implemented acquisition chain provides magnitude-only
received-power measurements, the objective is power-domain angular characterization and
repeatability assessment rather than transformation-based processing. The selected
sampling grid is therefore evaluated empirically through repeat-scan consistency and
correlation with simulation-derived power-domain reference trends.

To ensure repeatable data acquisition, the system employs deterministic sequencing of
robot pose execution and RF readout. At each sampling location, the manipulator
executes the target pose and undergoes a brief quasi-static stabilization period
before the RF measurement is triggered. Time-aligned acquisition following quasi-static
pose stabilization ensures that each received-power sample is associated with a
well-defined spatial configuration, while avoiding any reliance on electromagnetic
phase coherence.

\subsection{Measurement Modality and Scope}

The proposed system performs amplitude-only measurements of received power and does not measure absolute or relative phase. In the present implementation, a
spectrum-analyzer-based receiver chain combined with a harmonic mixer is used to
acquire an angularly indexed received-power readout at the operating frequency as the
probe is positioned over a geometry-calibrated hemispherical sampling surface. Robotic
motion is used solely as a means of achieving controlled and repeatable spatial
positioning during mmWave received-power measurements.

At each probe position, the recorded measurement corresponds to the received power,
\begin{equation}
P_{\mathrm{rx}}(\theta,\phi,r) \propto \left| E_{\mathrm{loc}}(\theta,\phi,r) \right|^{2},
\label{eq:power_measurement}
\end{equation}
where $E_{\mathrm{loc}}(\theta,\phi,r)$ denotes the radiated electric field at the
probe location. Only the magnitude-squared quantity is observed experimentally, and
no phase information is available or inferred. All analysis and validation in this
work are therefore carried out exclusively in the power domain.

The acquired data are interpreted as angularly indexed received-power measurements and are evaluated using power-domain metrics such as repeatability, spatial consistency of the executed sampling surface, and correlation of angular trends with idealized free-space references derived from full-wave simulation. Absolute field quantities, vector field components, and probe-corrected field values are not measured.

The methodology is intended for engineering validation and comparative characterization of embedded mmWave transmitters and is not designed for classical near/far-field antenna metrology. In particular, the system does not perform coherent signal acquisition, probe-corrected field processing, or transformation-based post-processing. Instead, the contribution of this work is a repeatable measurement methodology based on
geometry-calibrated spatial sampling and time-aligned pose-indexed received-power
acquisition, enabling portable characterization under realistic installation
constraints.

Dominant contributors to measurement uncertainty include robot positional repeatability, bracket orientation tolerance, probe polarization alignment, receiver chain stability, environmental multipath, and DUT output variability. These factors can affect both the measured received-power level and the apparent angular trend, particularly in non-anechoic environments and for compact embedded modules where installation details strongly influence the measured response. While a full quantitative uncertainty budget is beyond the scope of this work, these effects are partially bounded through repeated measurements and day-to-day repeatability analysis. The reported results therefore emphasize repeatability, spatial consistency of the executed sampling surface, and power-domain correlation against idealized references derived from full-wave simulation rather than absolute metrological accuracy.

\subsection{Full–Wave Electromagnetic Simulation Reference}

\begin{figure}[t!]
\centerline{\includegraphics[width=0.5\textwidth]{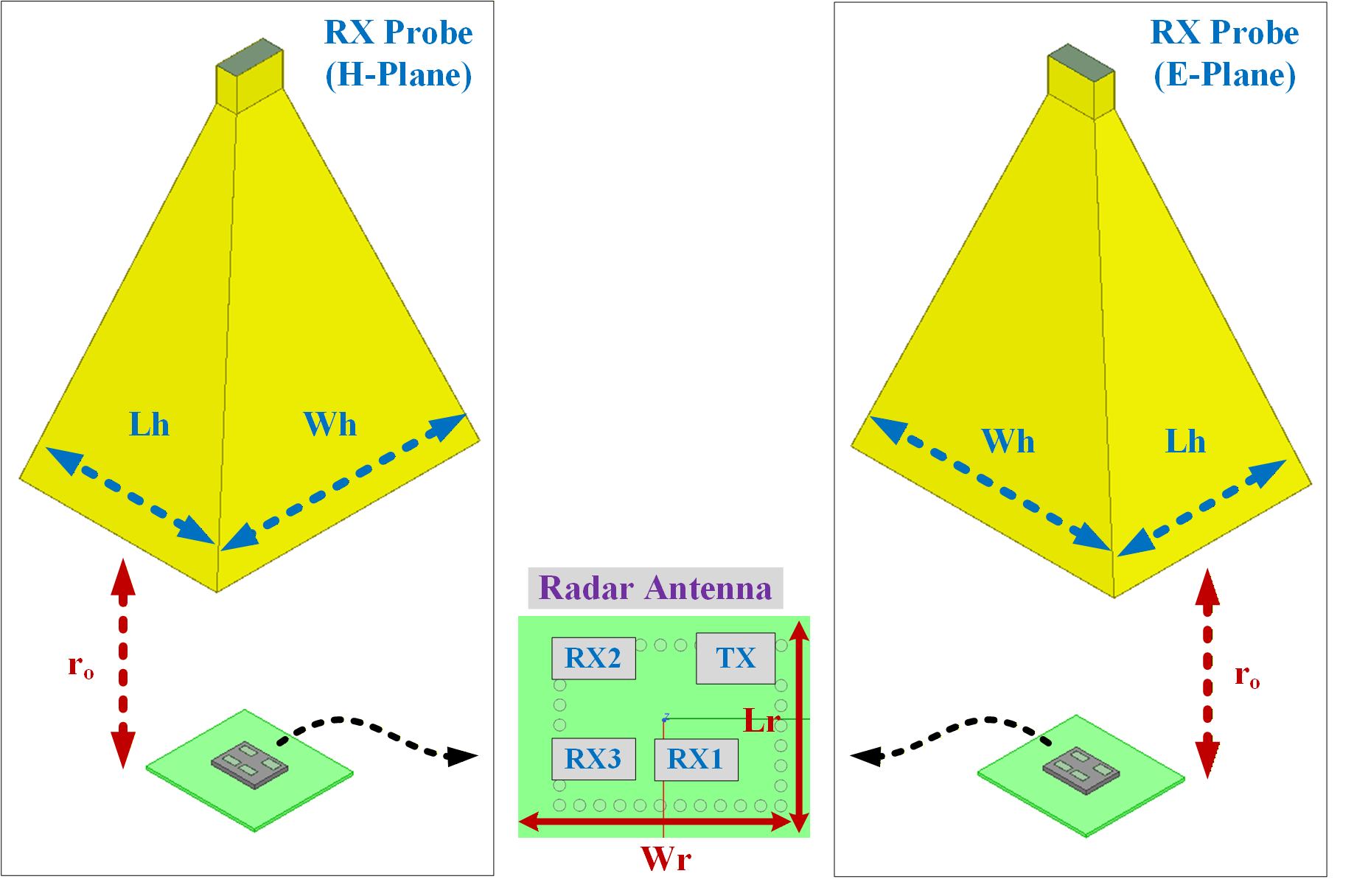}}
   \caption{Full-wave simulation setup illustrating the embedded mmWave radar module layout, including the transmit (Tx) antenna and multiple receive (Rx) channels, together with the receive-probe orientation in the H-plane and E-plane configurations.}
   \label{fig4}  
\end{figure}

To establish an idealized free-space reference for validating the proposed measurement methodology, a full-wave finite-element model of the receive (Rx) horn probe and the embedded radar module was implemented using a commercial electromagnetic solver. The simulation setup, illustrated in Fig.~\ref{fig4}, provides a baseline reference response under reflection-minimized boundary conditions. This idealized configuration supports controlled power-domain comparison by separating installation-dependent measurement effects from the modeled free-space behavior.

In the simulation model, the Rx horn probe is recreated with its measured geometric
parameters, including the flare profile, aperture width $W_h = 40~\mathrm{mm}$, and
aperture length $L_h = 30~\mathrm{mm}$. The horn is positioned at a separation distance
$r_0$ above the radar module, and two orthogonal probe orientations are considered:
an H-plane configuration, in which the aperture is aligned such that the dominant field
variation is along $L_h$, and an E-plane configuration, in which the horn is rotated by
$90^\circ$ such that the dominant field variation is along $W_h$, as depicted in the
left and right panels of Fig.~\ref{fig4}, respectively. In the physical measurement
setup, these two orientations are realized by rotating the horn by $90^\circ$ at the
end-effector, enabling consistent acquisition of both H- and E-plane cuts using the
same hemispherical sampling framework.

The embedded mmWave radar module is modeled with its transmit (Tx) and receive
(Rx1--Rx3) antenna elements and overall antenna footprint dimensions
$W_r = 6.5~\mathrm{mm}$ and $L_r = 5~\mathrm{mm}$. In the experimental measurements,
the Tx element of the module acts as the transmitter, while the scanning horn serves as
the external Rx probe. For calibration and repeatable measurements, the received signal
is monitored on the RX3 channel of the radar, providing a consistent Tx--Rx link
configuration across experiments. Absorbing boundary conditions and a surrounding
radiation region are employed to emulate free-space operation.

To generate a spherical reference dataset, the simulation domain is sampled over a
discrete grid defined by radial distance $r$, azimuthal angle $\phi$, and polar angle
$\theta$. The sampling coordinates follow the hemispherical geometry in Fig.~\ref{fig3}
and are mapped to Cartesian space using the standard spherical transformation: $x = r \sin\theta \cos\phi$, $y = r \sin\theta \sin\phi$, and $z = r \cos\theta$, where $\theta \in [\theta_{\min}, \theta_{\max}]$ and $\phi \in [\phi_{\min}, \phi_{\max}]$ denote the angular coverage used in the hemispherical scans.

For consistency with the robotic measurements, the full-wave solution is evaluated at (i) a fixed radius $r = r_0$ corresponding to the physical horn--to--DUT separation, (ii) an azimuthal span $\phi \in [\phi_1,\phi_2]$ matching the executed robotic trajectory, and (iii) a polar span $\theta \in [\theta_1,\theta_2]$ corresponding to the accessible hemisphere. Although the solver provides access to complete complex field quantities in the simulation domain, the present work performs validation using magnitude-only reference quantities to remain consistent with the implemented measurement modality.

Accordingly, the simulation results are reduced to a magnitude-based angular reference that represents the effective transmit--receive coupling between the radar Tx antenna and the horn probe at the measurement radius. This link-level reference captures the combined effects of radiation, propagation, and reception in a form that is directly comparable to the received-power measurements obtained experimentally. Comparisons between simulation and measurement are therefore carried out in the power domain and interpreted primarily in terms of angular trends rather than absolute link magnitude.

Discrepancies between measurement and simulation are expected due to environmental scattering, absorber limitations, robotic structure interactions, and cabling effects that are not included in the full-wave model. The objective of the comparison is not absolute agreement, but assessment of angular trend correlation and repeatability under realistic measurement conditions.

Throughout this paper, a strict distinction is maintained between measured quantities and simulation-derived reference quantities. The experimentally measured quantity is the received power $P_{rx}(\theta,\phi,r)$ obtained from the spectrum-analyzer-based receiver chain and evaluated exclusively in the power domain. The simulation provides a reference field magnitude $|E_{\mathrm{ref}}(\theta,\phi,r)|$ evaluated under the same geometric conditions; this quantity is shown only as a simulation-domain reference and is not directly measured by the proposed system. No experimentally measured phase or complex field quantities are used.

\begin{figure}[t!]
    \centering
    \includegraphics[width=0.5\textwidth]{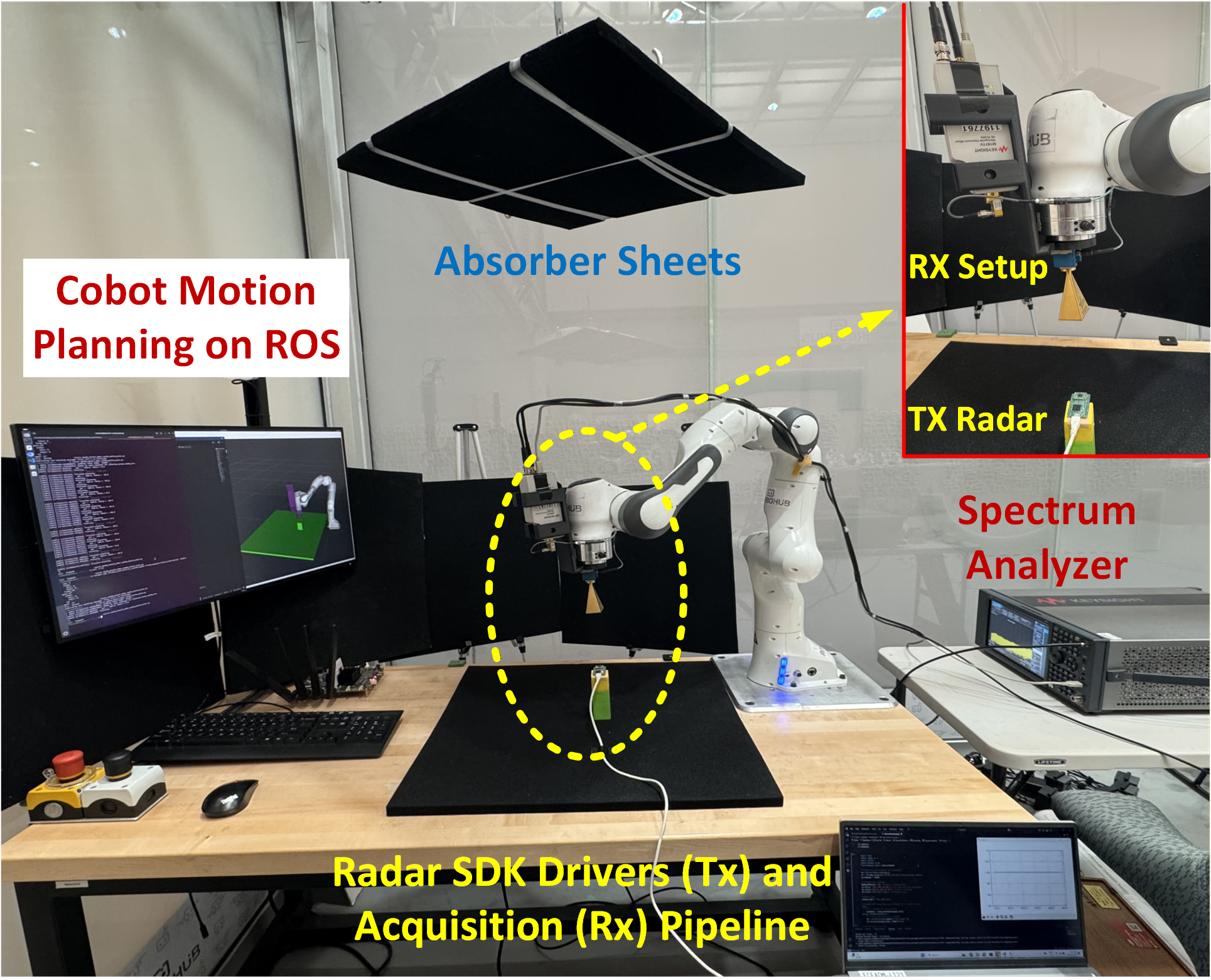}
    \caption{Experimental measurement setup showing the embedded radar DUT, collaborative robotic manipulator with horn receive probe, spectrum analyzer, radar control and time-aligned data acquisition systems, and overhead RF absorber sheets used to mitigate environmental multipath during received-power measurements.}
    \label{fig5}
    \vspace{-2mm}
\end{figure}

\subsection{Measurement Environment and Hardware Configuration}

The proposed autonomous measurement system operates within a semi-anechoic,
absorber-augmented workspace designed to suppress dominant multipath reflections while
maintaining flexibility for in-situ testing. As illustrated in Fig.~\ref{fig5}, the
device under test (DUT) is mounted at a fixed reference location on the table, while a
collaborative robotic manipulator positions the horn probe along the planned
hemispherical sampling grid. Overhead and table-mounted absorber sheets are arranged to
preserve a clear measurement region around the DUT and to attenuate reflections from
the ceiling, walls, and table surface that would otherwise distort angular received-power measurements. A spectrum analyzer, controlled through the measurement software, records received power in a time-aligned manner with the executed robot poses, enabling fully unattended scan execution once a campaign is configured. The monitor in Fig.~\ref{fig5} displays the RViz planning environment, including the collision geometry models used to ensure that trajectories remain collision-free while respecting the calibrated DUT--probe geometry.

A 7-DoF Franka Emika Panda collaborative robot serves as the primary manipulator for spatially positioning the receiving antenna throughout the measurement volume. Consistent with the workflow in Fig.~\ref{fig2} and the hemispherical sampling geometry in Fig.~\ref{fig3}, the robot’s high repeatability, integrated torque sensing, and smooth trajectory execution enable stable and repeatable probe placement across the
sampling domain. A custom 3D-printed end-effector bracket mounts the receiving horn antenna and harmonic mixer, positioning the RF front end at a known and fixed geometric offset relative to the robot flange. This configuration maintains a consistent probe reference point within the system’s coordinate hierarchy, enabling repeatable spatial alignment, collision-free hemispherical motion, and measurements at multiple scanning radii without the infrastructure demands of a fully anechoic chamber. The resulting modular hardware arrangement supports adaptation to varying DUT geometries and angular coverage requirements.

The measurement platform incorporates the surrounding table geometry and DUT fixture
into the robot planning environment as static collision volumes, preventing unmodeled
contact with the workspace. The DUT may be placed directly on the table surface or
elevated using a calibrated riser, depending on the desired angular coverage. Direct
placement limits the achievable polar angles due to mechanical interference, typically
restricting coverage to approximately $\pm 56^\circ$. Elevating the DUT by \SI{0.1}{m}
expands the reachable hemispherical region to nearly $\pm 70^\circ$.

The Panda robot provides seven revolute joints enabling dexterous maneuvering around
the DUT. One operational constraint arises from the $\pm 180^{\circ}$ rotation limit of
the final wrist joint, which can introduce local infeasibility near aggressive polar
motions. The scanning framework therefore incorporates joint-limit-aware motion planning
to avoid wrist singularities and to maintain smooth transitions across large angular
spans, substantially improving angular coverage for power-domain angular received-power characterization.

\subsection{Hemispherical Sampling, Calibration, and Data Acquisition Framework}

Accurate calibration of the coordinate transformations between the robot base, flange, custom end-effector bracket, and the probe reference point is critical for ensuring that each commanded spherical sampling pose corresponds precisely to a physical antenna position and orientation in space.

The robot base frame is defined using a predefined home configuration $\{\theta_i^{*}\}_{i=1}^{7}$, which serves as a reference posture for calibration. Applying forward kinematics (FK) and inverse kinematics (IK) to this configuration yields the base-to-flange transformation,
\begin{equation}\label{eq:1}
\mathbf{T}_{\text{base}} = \text{FK/IK}(\{\theta_i^*\}), \quad \mathbf{T}_{\text{base}} \in SE(3)
\end{equation}
which anchors all subsequent transformations within the robot-centric coordinate system.

The custom-designed horn antenna bracket introduces fixed rotational and translational offsets relative to the robot flange. The bracket provides a nominal yaw rotation $\gamma \approx -100^\circ$, minimal pitch and roll angles ($\beta, \alpha \approx 0^\circ$), and a small translational displacement $\mathbf{p}_{\text{offset}} = (x_o, y_o, z_o)$. Precision dial-gauge measurements and manufacturer specifications yield translational accuracies of approximately $\pm 2$\,mm and rotational accuracies of $\pm 1^\circ$, which are sufficient to maintain antenna orientation integrity during scanning. The corresponding homogeneous transformation is given by,
\begin{equation} \label{eq:2}
\mathbf{T}_{\text{offset}} =
\begin{bmatrix}
    \mathbf{R}_{\text{offset}}(\gamma,\beta,\alpha) & \mathbf{p}_{\text{offset}}(x_o,y_o,z_o) \\
    \mathbf{0}^\top & 1
\end{bmatrix}
\end{equation}
where $\mathbf{R}_{\text{offset}}$ denotes the composite rotation derived from sequential Euler rotations. This calibrated offset ensures that commanded scanning poses align accurately with the receiving antenna’s physical probe reference point. The orientation offset is constructed using a Z--Y--X intrinsic Euler rotation sequence, $R_{\text{offset}}(\gamma,\beta,\alpha) = R_z(\gamma) R_y(\beta) R_x(\alpha)$.

\begin{table}[t!]
    \centering
    \caption{Representative Parameter Values for the Experimental Setup}
    \label{table:1}
    \resizebox{\linewidth}{!}{
        \begin{tabular}{>{\bfseries}c c c}
            \toprule
            \rowcolor{gray!30}\textbf{Configuration Parameter} & \textbf{Role} & \textbf{Typical Values} \\
            \midrule
            $\gamma, \beta, \alpha$ & End-effector Euler offsets & $\gamma \approx -100^\circ$, $\beta \approx 0^\circ$, $\alpha \approx 0^\circ$ \\
            $x_o,y_o,z_o$ & Translational offsets & 0.02--0.05\,m \\
            $r$ & Spherical sampling radius & 0.03--0.20\,m \\
            $\phi$ & Azimuth sampling angles & $0^\circ$--$359^\circ$ \\
            $\theta$ & Polar sampling angles & $0^\circ$--$70^\circ$ \\
            $\nu_{\mathrm{max}}$ & Velocity scaling factor & 5--10\% \\
            $t_{\mathrm{dwell}}$ & Dwell time per pose & 1--20\,s \\
            \bottomrule
        \end{tabular}
    }
\end{table}

The proposed measurement framework samples received power over a structured spherical grid centered on the DUT. The grid is defined through two discrete angular sets, \[
\Phi = \{\phi_i \mid 0^{\circ} \le \phi_i < 360^{\circ}\}, \\
\Theta = \{\theta_j \mid -\theta_{\max} \le \theta_j \le \theta_{\max}\},
\]
with $\theta_{\max}$ typically between $56^{\circ}$ and $70^{\circ}$ depending on workspace constraints. In the implemented hemispherical scans, $\theta$ is sampled over $[0,\theta_{\max}]$, corresponding to the upper hemisphere with the probe oriented toward the DUT. The symmetric notation is used solely to describe a general elevation span; in practice, measurements are confined to the accessible half-space dictated by the robot workspace and probe mounting constraints. The planning framework supports angular resolutions down to $2.5^{\circ}$ for dense hemispherical coverage. In the experimental configurations of Section~IV, angular step sizes of $10^{\circ}$–$20^{\circ}$ are employed as a practical trade-off between measurement density and total scan duration. These increments provide sufficient angular resolution for comparative, magnitude-only validation, enabling assessment of angular agreement with simulation-derived power-domain references and evaluation of measurement repeatability under realistic installation conditions.

Each pair $(\phi,\theta)$ corresponds to a unique spherical direction from the DUT. The system supports multiple radii $r \in \{0.03\text{m} \text{--} 0.20\text{m}\}$. Measurements are performed at finite distances that exhibit increasingly far-field--like angular power behavior as the measurement radius increases. Here, far-field-like refers to stabilization of angular received-power distributions rather than satisfaction of formal Fraunhofer criteria. In this work, multiple probe radii are evaluated to quantify how the measured angular received-power response evolves with distance and to identify radii at which the angular distributions become approximately invariant to further increases in separation. Smaller radii provide finer spatial resolution at the cost of increased collision sensitivity, whereas larger radii reduce collision risks but require longer scanning times. The spherical sampling transformation is expressed as:
\begin{equation}\label{eq:3}
\mathbf{T}_{\text{sphere}} = 
\begin{bmatrix}
    R_z(\phi) R_y(-\theta) &
    r 
    \begin{bmatrix}
        \sin \theta \cos \phi \\
        \sin \theta \sin \phi \\
        \cos \theta
    \end{bmatrix} \\
    \mathbf{0}^\top & 1
\end{bmatrix}
\end{equation}
where the negative rotation about the $y$-axis accounts for the upward-facing orientation of the robot flange relative to the global $z$-axis.

At each feasible pose, the system incorporates a dwell period, $t_{\mathrm{dwell}}$, of 1-20 seconds to attenuate mechanical vibrations and ensure measurement stability. The receiving chain, consisting of a horn antenna, harmonic mixer, and spectrum analyzer, records received power or S-parameters using standard SCPI commands. All measurements are indexed by $(\phi,\theta,r)$ and stored in structured logs. Typical scans contain 100--300 valid poses and may run from several minutes to multiple hours depending on angular resolution, dwell time, and planning success rates. During motion, the robot
operates with a conservative joint-velocity scaling factor $\nu_{\mathrm{max}}$ (5–10\% of the nominal joint speed), which enforces quasi-static trajectories suitable for mmWave measurements. Table~\ref{table:1} summarizes the standard heuristic parameters used in the presented methodology, including bracket orientation, translational offsets, sampling radii, angular steps, and motion- and measurement-related settings. The parameter ranges in Table~\ref{table:1} reflect practical limits imposed by (i) collision-aware reachability, (ii) scan-time constraints dominated by dwell and instrument settling, and (iii) maintaining quasi-static probe placement for repeatable power-domain sampling.


\begin{table*}[t!]
\caption{Transformation Components Specification}
\centering
\label{table:2}
\begin{tabular}{cccccc}
\toprule 
\rowcolor{gray!30}
\textbf{Transform} & \textbf{Role} & \textbf{Key Variables} & \textbf{Derivation / Code} & \textbf{Matrix Form} \\ \midrule
$\mathbf{T}_{\mathrm{base}}$ (Eq.~\ref{eq:1}) & 
\begin{tabular}[c]{@{}c@{}}Anchors to global coordinates\end{tabular} & 
\begin{tabular}[c]{@{}c@{}}
$\mathbf{R}_{\mathrm{base}}$: orientation matrix \\
$\mathbf{p}_{\mathrm{base}}$: position vector
\end{tabular} & 
\texttt{get\_FK/IK()} service call & 
$\begin{bmatrix}
\mathbf{R}_{\mathrm{base}} & \mathbf{p}_{\mathrm{base}} \\
\mathbf{0}^\top & 1
\end{bmatrix}$ \\ \\

$\mathbf{T}_{\mathrm{offset}}$ (Eq.~\ref{eq:2}) & 
\begin{tabular}[c]{@{}c@{}}Encodes end-effector orientation\\ plus translation offsets\end{tabular} & 
\begin{tabular}[c]{@{}c@{}}
$\gamma,\beta,\alpha$: Euler angles \\
$x_o,y_o,z_o$: offset distances
\end{tabular} & 
\texttt{get\_wpose()} & 
$\begin{bmatrix}
\mathbf{R}_{\mathrm{offset}} & \mathbf{p}_{\mathrm{offset}} \\
\mathbf{0}^\top & 1
\end{bmatrix}$ \\ \\

$\mathbf{T}_{\mathrm{sphere}}$ (Eq.~\ref{eq:3}) & 
\begin{tabular}[c]{@{}c@{}} Converts spherical to Cartesian \\ coordinates (Maps $(\phi,\theta,r)$) \end{tabular} & 
\begin{tabular}[c]{@{}c@{}}
$\phi$: azimuth, 
$\theta$: polar,  
$r$: radius \\
$\mathbf{R}_{\mathrm{sphere}}$: Rotation matrix
\end{tabular} & 
\texttt{get\_A\_T\_G(phi,theta,r)} & 
$\begin{bmatrix}
\mathbf{R}_{\mathrm{sphere}} & \mathbf{p}_{\mathrm{sphere}} \\
\mathbf{0}^\top & 1
\end{bmatrix}$ \\ \\

$\mathbf{T}_{\mathrm{final}}$ (Eq.~\ref{eq:Tfinal}) & 
\begin{tabular}[c]{@{}c@{}}Executable end-effector pose\end{tabular} & 
\begin{tabular}[c]{@{}c@{}}
Composite transformation: \\
$\mathbf{T}_{\mathrm{base}}$, $\mathbf{T}_{\mathrm{sphere}}$, $\mathbf{T}_{\mathrm{offset}}$
\end{tabular} & 
\texttt{set\_pose\_target()} & 
$\mathbf{T}_{\mathrm{base}} \times \mathbf{T}_{\mathrm{sphere}} \times \mathbf{T}_{\mathrm{offset}}$ \\ \hline
\end{tabular}
\end{table*}


\section{Motion Planning and Acquisition Pipeline}
\label{sec:III}

This section details the automation pipeline that links the hemispherical sampling model, robotic trajectory generation, and RF measurement subsystem into a unified end-to-end framework. Building on the coordinate transformations and environmental modeling presented in this Section, the proposed pipeline converts the abstract directional grid into physically realizable transmitter poses and synchronizes each robot motion with high-fidelity RF data acquisition. The integration of kinematics, collision-aware planning, and automated measurement control enables repeatable and scalable characterization of mmWave devices without human supervision.

\subsection{Final Pose Computation}
\label{subsec:VIA}

For each measurement point $(\phi,\theta,r)$ in the hemispherical sampling set, the
system computes an executable end-effector pose, denoted $\mathbf{T}_{\mathrm{final}}$.
This pose specifies the physical position and orientation of the receiving probe
relative to the DUT and must satisfy both the geometry-calibrated
sampling requirements and the robot’s kinematic feasibility constraints.

The computation follows the structured transformation hierarchy established in
Section~\ref{sec:III}. First, the global reference frame is defined by the base transform $\mathbf{T}_{\mathrm{base}}$, which anchors the robot to the laboratory coordinate system. Next, the desired sampling direction and distance are mapped to Cartesian space through the spherical transformation $\mathbf{T}_{\mathrm{sphere}}$, capturing the specified azimuth, polar angle, and radius. Finally, the calibrated end-effector offset $\mathbf{T}_{\mathrm{offset}}$ aligns the robot flange with the probe reference point. The resulting executable pose is obtained by composing these transforms as defined in \eqref{eq:Tfinal}.

This transformation chain establishes a direct mapping between the theoretical
hemispherical sampling grid and the physical placement of the receiving probe in
three-dimensional space. It ensures that (i) the probe orientation is aligned with the intended sampling direction, (ii) the effective measurement radius is preserved, and (iii) positional repeatability remains within the $\pm 2$~mm tolerance required for repeatable mmWave angular received-power measurements. For the 3--15~cm measurement radii used in this work, a $\pm 2$~mm deviation corresponds to approximately
$1.3\%$--$6.7\%$ relative variation in $r$, which, under the spherical $1/r$ amplitude
dependence, produces less than $0.6$~dB variation in received power magnitude over the
entire range. At 60~GHz (wavelength $\lambda \approx 5$~mm), this level of geometric
uncertainty is therefore small compared to the overall amplitude error budget reported
in Section~V and is consistent with practical positioning tolerances for repeatable power-domain mmWave measurements.

Each computed $\mathbf{T}_{\mathrm{final}}$ is verified for numerical stability and subsequently submitted to the motion planner, which resolves whether the desired pose is kinematically reachable and collision-free. Table~\ref{table:2} summarizes the individual transformation modules and their physical meaning. Function names shown in Table~\ref{table:2} refer to the open-source software implementation (ROS/MoveIt service calls); the underlying transformations are standard SE(3) homogeneous transforms.

\subsection{Collision-Aware Motion Planning}

For each sampling direction $(\phi,\theta,r)$, the robotic system evaluates whether the
corresponding end-effector pose $\mathbf{T}_{\mathrm{final}}$, generated through the
transformation chain described in the previous section, can be executed safely and
accurately by the manipulator. Because the mmWave received-power measurements performed
in this work are sensitive to probe positioning and orientation repeatability, the
motion planner enforces quasi-static trajectories with conservative velocity and
acceleration bounds. These constraints reduce dynamic disturbances at the receiving
probe and help maintain stable spatial alignment relative to the DUT during measurement acquisition.

The planning pipeline begins with a feasibility analysis in configuration space. First, the system evaluates whether an inverse-kinematics (IK) solution exists within the robot’s joint limits. Because the hemispherical sampling grid drives the robot into high-curvature regions of its workspace, the feasibility check also evaluates the conditioning of each IK solution by examining the Jacobian determinant. This step filters out wrist configurations that lie close to singularities or exhibit poor manipulability, which are known to cause instability in end-effector orientation and may jeopardize measurement integrity.

Each candidate solution that satisfies the singularity criteria undergoes collision pre-validation using a detailed planning-scene representation that includes the robot links, the custom bracket, the DUT, and the support table. Only configurations that remain strictly within the collision-free set are considered admissible. For these valid configurations, the planner selects the one farthest from the collision boundary to maximize robustness during trajectory optimization.

\begin{algorithm}[t!]
\caption{Collision-Aware Cobot Motion Planning}
\label{alg:motion_planning}
\SetAlgoLined
\DontPrintSemicolon

\KwIn{Desired end-effector pose $\mathbf{T}_{\mathrm{final}} \in SE(3)$, current configuration $\mathbf{q}_{\mathrm{start}}$, home configuration $\mathbf{q}_{\mathrm{home}}$}
\KwOut{Feasible trajectory $\tau$ or failure}

$\mathcal{C}_{\mathrm{joint}} \gets \{\mathbf{q} : q_i^{\min} \le q_i \le q_i^{\max}\}$\;
$\mathcal{C}_{\mathrm{free}} \gets \{\mathbf{q} \in \mathcal{C}_{\mathrm{joint}} : \neg \mathrm{Collision}(\mathbf{q})\}$\;

$\mathcal{Q}_{\mathrm{IK}} \gets \mathrm{IK\_Solve}(\mathbf{T}_{\mathrm{final}})$\;
$\mathcal{Q}_{\mathrm{IK}} \gets \{\mathbf{q} \in \mathcal{Q}_{\mathrm{IK}} : \mathbf{q} \in \mathcal{C}_{\mathrm{joint}}\}$\;
\If{$\mathcal{Q}_{\mathrm{IK}} = \emptyset$}{
    \Return failure\;
}

$\mathcal{Q}_{\mathrm{valid}} \gets \emptyset$\;
\ForEach{$\mathbf{q} \in \mathcal{Q}_{\mathrm{IK}}$}{
    \If{$\sigma_{\min}(J(\mathbf{q})) > \epsilon_{\mathrm{sing}}$ \textbf{and} $\mathbf{q} \in \mathcal{C}_{\mathrm{free}}$}{
        $\mathcal{Q}_{\mathrm{valid}} \gets \mathcal{Q}_{\mathrm{valid}} \cup \{\mathbf{q}\}$\;
    }
}
\If{$\mathcal{Q}_{\mathrm{valid}} = \emptyset$}{
    \Return failure\;
}

$\mathbf{q}^\star \gets \arg\min_{\mathbf{q} \in \mathcal{Q}_{\mathrm{valid}}} \|\mathbf{q}-\mathbf{q}_{\mathrm{start}}\|_2$\;

$\tau \gets \mathrm{OMPL\_Plan}(\mathbf{q}_{\mathrm{start}}, \mathbf{q}^\star; s_{\mathrm{vel}}, s_{\mathrm{acc}})$\;
\If{$\tau \neq \emptyset$}{
    \Return $\tau$\;
}

$\tau_{\mathrm{home}} \gets \mathrm{OMPL\_Plan}(\mathbf{q}_{\mathrm{start}}, \mathbf{q}_{\mathrm{home}}; s_{\mathrm{vel}}, s_{\mathrm{acc}})$\;
\If{$\tau_{\mathrm{home}} = \emptyset$}{
    \Return failure\;
}

\textbf{move to} $\mathbf{q}_{\mathrm{home}}$; $\mathbf{q}_{\mathrm{start}} \gets \mathbf{q}_{\mathrm{home}}$\;

$\tau \gets \mathrm{OMPL\_Plan}(\mathbf{q}_{\mathrm{start}}, \mathbf{q}^\star; s_{\mathrm{vel}}, s_{\mathrm{acc}})$\;
\If{$\tau \neq \emptyset$}{
    \Return $\tau$\;
}

\Return failure\;
\end{algorithm}

Open Motion Planning Library (OMPL) is then invoked to compute a dynamically feasible trajectory from the current joint configuration to the selected target configuration. The planner uses a reduced velocity and acceleration scaling (5\% of nominal) to maintain quasi-static behavior. If the primary planning attempt fails, typically due to joint-limit proximity, geometric discontinuities, or local minima in the configuration space the system initiates a recovery routine. In this routine, the robot first returns to a predefined home configuration, which provides a well-conditioned starting point with improved manipulability. From this configuration, the planner reattempts trajectory generation toward $\mathbf{T}_{\mathrm{final}}$. If this second attempt fails, the pose is declared unreachable. This hierarchical strategy consisting of IK feasibility filtering, singularity rejection, collision reasoning, and multi-stage recovery provides both robustness and reproducibility during long-duration hemispherical scans. The complete mathematical structure of the planning logic is summarized in Algorithm~\ref{alg:motion_planning}.

This motion-planning architecture provides a structured and repeatable mechanism for navigating the kinematic and geometric constraints of the measurement environment. By integrating feasibility checks, explicit singularity avoidance, and deterministic recovery procedures, the planner ensures that each pose in the hemispherical sampling grid is executed with the positional stability and orientation consistency required for repeatable mmWave received-power measurements. Robotic motion is used solely to achieve controlled and repeatable spatial positioning, rather than to perform any electromagnetic processing. The effectiveness of this collision-aware,
Jacobian-conditioned planning strategy is quantified in the following section.
For the scan configurations used in this work, the selected RRT-Connect planner
achieved a 100\% planning success rate with short planning times and highly repeatable
trajectories.



\subsection{Automated RF Measurement Pipeline}
\label{subsec:automation}

Following successful execution of the robot trajectory and stabilization of the end-effector at the desired pose $T_{\text{final}}$, the system initiates a fully automated receive–transmit (Rx–Tx) interrogation of the DUT. Because mmWave measurements are highly sensitive to mechanical perturbations, even at the sub-millimeter scale, the pipeline enforces a mandatory dwell interval at each sampling point to ensure that all residual vibrations have dissipated. This guarantees that the measured power corresponds strictly to the DUT’s angular radiation characteristics and is not corrupted by transient motion of the receiving horn.

The RF subsystem communicates with the Keysight N9040B PXA signal analyzer via PyVISA using a high-throughput HS-LINK interface. A WR-15 harmonic mixer is used as a mmwave front-end, providing the waveguide interface to the horn probe and downconverting the 58--63~GHz radar band to an intermediate frequency that lies within the input range of the PXA while preserving high dynamic range in the receive chain. Upon initialization, the spectrum analyzer is configured with a fixed center frequency, frequency span, resolution bandwidth, video bandwidth, and MAX-HOLD detector mode to ensure consistent acquisition conditions across all measurements. These parameters are selected such that the analyzer noise floor remains well below the weakest expected response from the device under test, while the harmonic-mixer–analyzer chain operates within its linear region, preserving the available dynamic range of the receive path. During each acquisition cycle, a new trace is triggered via SCPI commands and the full
spectral vector is retrieved. The received power at the operating frequency is then
extracted from the trace with 16-bit digital resolution and associated with the
corresponding spatial sampling pose.

To maintain measurement integrity, each extracted value undergoes a numerical validity check, comparing it against bounds derived from the link budget, noise floor, and expected dynamic range. Valid measurements are committed to disk using atomic write operations, which prevent data corruption during unexpected interruptions or system faults. Each entry is indexed using its corresponding spherical coordinates $(\phi,\theta,r)$ and a timestamp, enabling robust power-domain angular characterization of the complete hemispherical. This screening ensures that only measurements acquired within the calibrated linear operating region of the receiver chain are retained for subsequent angular power-pattern mapping and quantitative comparison against full-wave reference predictions. Algorithm~\ref{alg:measurement_pipeline} summarizes the complete RF acquisition protocol.

\begin{algorithm}[t!]
\caption{Automated RF Measurement Pipeline}
\label{alg:measurement_pipeline}
\SetAlgoLined
\DontPrintSemicolon

\textbf{Initialization:}\;
$\triangleright$ Establish instrument control session (PyVISA)\;
$\triangleright$ Configure spectrum analyzer: $f_c$, span, RBW/VBW, detector = MAX-HOLD\;
$\triangleright$ Preallocate data file; write metadata and headers\;

\BlankLine
\textbf{Measurement Loop:}\;
\For{each sampling point $(\phi,\theta,r)$}{
$\triangleright$ Reset trace state and trigger a new acquisition (SCPI)\;
$\triangleright$ Enforce quasi-static stabilization dwell $t_{\mathrm{dwell}}$\;
$\triangleright$ Retrieve spectral trace vector\;
$\triangleright$ Extract received power $P_{\mathrm{rx}}$ at $f_c$ (16-bit digital resolution)\;
$\triangleright$ Apply range/quality checks on $P_{\mathrm{rx}}$\;
$\triangleright$ Atomically append $(\phi,\theta,r,P_{\mathrm{rx}},t)$ to file\;
$\triangleright$ Log acquisition diagnostics (timestamp, index)\;
}

\BlankLine
\textbf{Termination:}\;
$\triangleright$ Close instrument session and release resources\;
$\triangleright$ Verify dataset completeness; flag missing or invalid entries\;

\end{algorithm}

The initialization phase configures the spectrum analyzer with fixed spectral parameters
prior to the start of a scan, ensuring consistent acquisition conditions and eliminating
intra-scan parameter drift. Within the measurement loop, a single acquisition is
performed for each spatial sampling point. Before each acquisition, the trace state is
reset and the analyzer is configured in MAX-HOLD mode, ensuring that the recorded trace
captures the maximum received power observed during the dwell interval rather than
accumulating residual content from previous poses. A dwell period is enforced at each pose to allow the manipulator to reach quasi-static mechanical equilibrium before RF acquisition is triggered. The subsequent data-fetch operation retrieves the spectral trace directly from the instrument buffer. The received power at the operating frequency is then extracted as a scalar quantity, providing a pose-indexed representation of the DUT response in the power domain. Basic range and quality checks are applied to identify corrupted or out-of-range samples. Each validated measurement is written atomically to persistent storage together with its associated
angular coordinates and timestamp, ensuring data integrity during long-duration
automated scans. Upon completion of the scan, the termination stage closes the instrument control session and verifies dataset completeness by checking for missing or invalid entries. This structured acquisition pipeline ensures repeatable, time-aligned received-power measurements that are directly compatible with the geometry-calibrated spatial sampling framework described in the preceding sections.

This automated RF acquisition pipeline, when combined with collision-aware motion planning, enables fully autonomous hemispherical spatial sampling for mmWave received-power measurements with high repeatability and controlled geometric accuracy. In the present implementation, the system measures received power magnitude at the operating frequency as a function of angular position and measurement radius. No absolute or relative phase information is acquired, and no complex-field quantities are measured or inferred. Extension to phase-coherent complex-field sampling would require different instrumentation, such as a vector network analyzer or coherent receiver, and is left for future work.


\section{Experiments}
\label{sec:V}
The baseline measurements represent a typical operator-dependent laboratory workflow and are included to assess repeatability and alignment consistency rather than absolute metrological accuracy. They provide a practical reference for evaluating the benefits of automated, geometry-calibrated spatial sampling relative to manual probe positioning.

This section presents an experimental evaluation of the proposed automated measurement platform, focusing on motion-planning reliability, calibration stability, received-power measurement repeatability, and correlation with simulation-derived power-domain references. All measurements were conducted in a semi-anechoic, absorber-augmented environment using the system architecture described in Section~\ref{sec:III}. The experimental workflow integrates collision-aware motion planning, automated RF acquisition, and transformation-based pose synthesis. A representative laboratory setup is illustrated in Fig.~\ref{fig5}.

\begin{figure*}[!t]
    \centering
    \subfloat[]{\includegraphics[width=\textwidth]{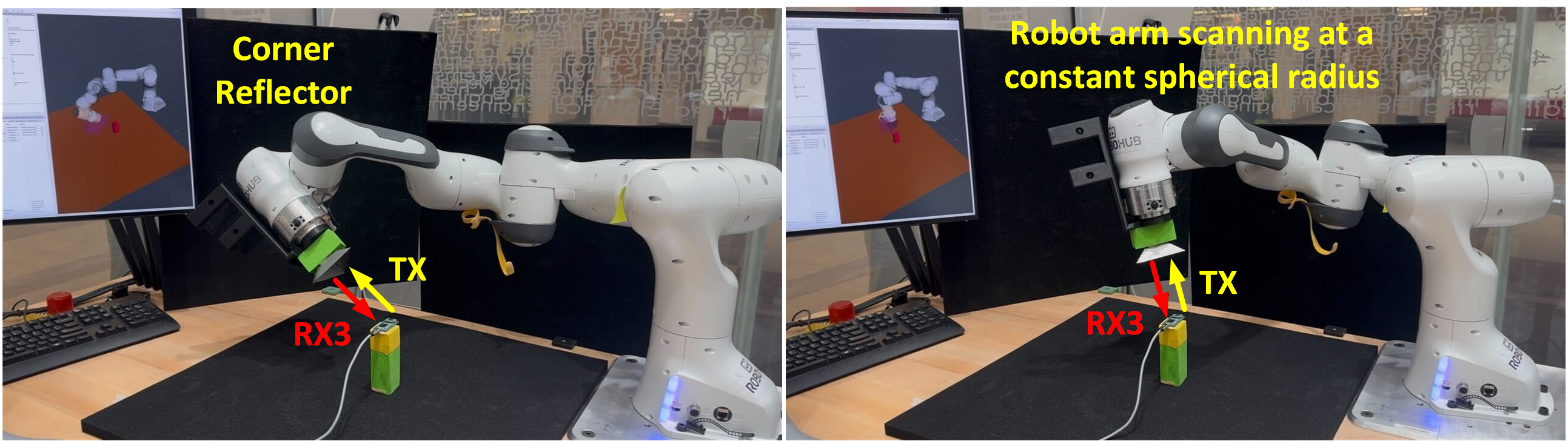}}
    \label{fig6a}
    \hfil
    \subfloat[]{\includegraphics[width=\textwidth]{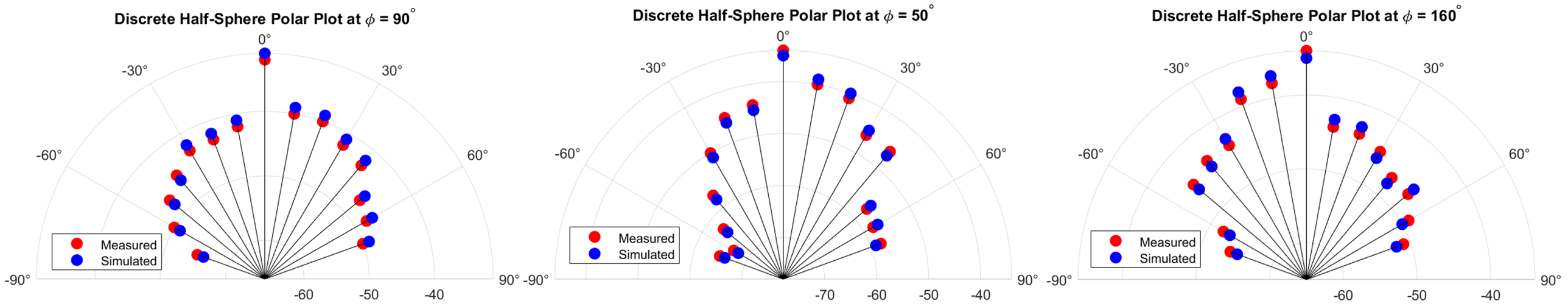}}
    \label{fig6b}
    \caption{Corner-reflector-based calibration and validation of the measurement system.
(a) Experimental setup showing the embedded mmWave radar module with the transmit
antenna and RX3 channel oriented toward a metallic trihedral corner reflector, while
the collaborative robot positions the horn receive probe along a hemispherical
trajectory at a fixed measurement radius. (b) Power-domain comparison between
simulation-derived reference results and measured received power from the corner
reflector over discrete hemispherical sampling at $\phi = 90^{\circ}$, $\phi = 50^{\circ}$, and $\phi = 160^{\circ}$.}
    \label{fig6}
\end{figure*}

To support reproducibility, experiments were conducted across multiple days with independent system resets and DUT re-mounting checks prior to each session. Before each scan, a metallic trihedral corner reflector was used as a link-level verification target to confirm receiver-chain stability and consistency of pose-to-angle association. As shown in Fig.~\ref{fig6}(a), the reflector is placed at a known distance and direction within the hemispherical workspace, and a short verification scan is performed to confirm that the measured peak occurs at the expected angular coordinates and that the received-power level remains consistent with a simple link-budget estimate for the transmit--receive chain. This procedure provides a practical consistency check for the geometric mapping and receive-path stability under the specific mounting configuration used for that session.

The consistency of this verification procedure is illustrated in Fig.~\ref{fig6}(b), which compares simulated and measured received power from the trihedral corner reflector over a discrete set of azimuth and elevation angles $(\phi,\theta)$ used in a representative verification scan. In this example, scans are performed in the $\phi = 50^{\circ}$, $90^{\circ}$, and $160^{\circ}$ planes, with elevation angles $\theta \in [-80^{\circ},80^{\circ}]$ sampled in $10^{\circ}$ steps at a constant probe distance $r = 5\,\mathrm{cm}$, as indicated in the figure. Agreement in the peak direction and the corresponding received-power trend supports the validity of the pose indexing and provides confidence that subsequent scans are acquired under consistent link conditions.


For intra-day robustness, repeated measurements were collected under identical conditions, while full power-cycle resets were performed between days to evaluate stability under realistic laboratory usage. In total, thousands of spatial samples were acquired across thirteen measurement radii spanning $r \in [\lambda/2,\,3\lambda]$ at 60~GHz. Measurements are therefore performed at finite distances that exhibit increasingly far-field--like angular received-power behavior as the radius increases.

All experiments were performed using (i) a Franka Emika Panda 7-DoF collaborative robot operating under a ROS/MoveIt framework, (ii) a Keysight N9030B PXA signal analyzer paired with an M1971V WR-15 harmonic mixer, and (iii) a Quinstar WR-15 horn antenna functioning as the receiving probe. The DUT remained stationary throughout each scan, while the horn probe followed hemispherical sampling trajectories defined by the transformation chain. A uniform angular sampling strategy was adopted to enable systematic comparison across radii and with simulation-derived power-domain references. For even radii (4, 6, 8, 10, 12, 14\,cm), scans used $\Delta\phi=\Delta\theta=15^\circ$, while for odd radii (3, 5, 7, 9, 11, 13\,cm), a coarser $20^\circ$ increment was employed. The 5\,cm configuration was additionally sampled at $10^\circ$ resolution to evaluate higher-density sampling. These schemes produced distinct $(\phi,\theta)$ grids containing 72--252 total points per radius.


\subsection{Motion Planning and Hemispherical Coverage}

The motion-planning subsystem is a key enabling component of the proposed measurement methodology, as it governs the robot’s ability to execute each commanded pose on the hemispherical sampling grid while respecting joint limits, collision constraints, and the quasi-static motion requirements of mmWave received-power measurements. To evaluate this component systematically, several widely used planning algorithms RRT-Connect, BIT*, PRM*, CHOMP, and a reinforcement-learning–based PPO planner were benchmarked under identical workspace and kinematic conditions. These algorithms span sampling-based, optimization-based, and learning-based paradigms, enabling a broad assessment of planning characteristics relevant to repeatable geometry-calibrated spatial sampling.

Each planner was evaluated in terms of planning time, trajectory length, and success
rate, providing insight into computational efficiency and reliability in the confined
measurement workspace. In addition, repeatability and calibration stability were
assessed using two independent Franka Panda manipulators (“Panda-1” and “Panda-2”) to
evaluate how consistently commanded poses are reproduced across hardware instances.
Calibration stability was quantified by monitoring the flange-to-bracket transformation
over multiple sessions, providing a practical measure of end-effector alignment consistency following power cycling and reinitialization.

Table~\ref{tab:3} summarizes these planning and calibration metrics. The planning-related values characterize how effectively each algorithm navigates the constrained workspace surrounding the DUT, particularly at small measurement radii where collision margins are limited. The calibration-related metrics quantify pose repeatability and alignment stability and are expressed in terms of received-power variability observed across repeated measurements under identical geometric conditions. Here, repeatability is computed as the mean absolute difference (MAE) of the measured received power across repeated scans acquired under identical geometric conditions, while calibration error denotes the measured session-to-session variation of the flange-to-bracket alignment estimate.

Among the tested planners, RRT-Connect provided the most favorable trade-off between planning time, trajectory length, and success rate, and was therefore selected for all subsequent measurement experiments. The remaining planners in Table~\ref{tab:3} are included to demonstrate that the proposed framework is agnostic to the choice of motion-planning backend and to contextualize the selected configuration. A detailed discussion of the numerical comparisons and their implications for scanning reliability is provided in the Discussion section.

All motion planners were benchmarked using the same set of target poses drawn from the $(\phi,\theta,r)$ sampling grids in Table~\ref{tab:IV}. To ensure a fair comparison, all runs employed an identical planning scene, including the robot model and collision geometry for the bracket, DUT, and support table, as well as identical kinematic constraints (joint limits and velocity-scaling settings). Planning success is defined as the fraction of commanded target poses for which the planner returned a collision-free trajectory with an inverse-kinematics-feasible goal configuration. The reported trajectory length corresponds to the cumulative joint-space path length of the planned trajectory, computed as the sum of absolute joint increments over all trajectory waypoints. Day-to-day repeatability is evaluated using the same power-domain MAE metric as intra-day repeatability, but computed across scans acquired on different days after full system re-initialization.

\begin{table*}[htbp]
\centering
\setlength{\tabcolsep}{5pt}  
\caption{Motion-planning benchmark and power-domain repeatability metrics (including calibration stability).}
\small
\begin{tabular}{cccccccc}
\toprule 
\rowcolor{gray!30}
\textbf{Algorithm} & \textbf{Category} & \textbf{Planning} & \textbf{Success} & \textbf{Traj. Length} & \textbf{Repeatability} & \textbf{Calib. Error} & \textbf{Day-to-Day} \\
\rowcolor{gray!30}
 & & \textbf{Time} (s) & \textbf{Rate} (\%) & (m) & \textbf{(Power MAE, dB)} & (mm) & \textbf{Repeat.} (dB) \\
\midrule
RRT-Connect \cite{orthey2023sampling} & Sampling-based & 0.45 & 100 & 3.14 & 0.18 & 0.37 & 0.24 \\
BIT* \cite{gammell2015batch} & Optimal Sampling & 1.50 & 100 & 2.98 & 0.21 & 0.35 & 0.20 \\
PRM* \cite{orthey2023sampling} & Sampling-based & 2.80 & 100 & 3.11 & 0.22 & 0.41 & 0.28 \\
CHOMP \cite{liu2022benchmarking} & Optimization & 3.20 & 98 & 2.92 & 0.15 & 0.36 & 0.22 \\
PPO (Deep RL) \cite{schulman2017proximal} & Learning-based & 0.62 & 100 & 3.13 & 0.19 & 0.39 & 0.19 \\
\bottomrule
\end{tabular}
\label{tab:3}
\end{table*}

To evaluate geometric robustness and operational consistency, multiple hemispherical scan configurations were executed by varying the sampling radius, angular resolution, and accessible polar range. Thirteen measurement radii spanning 3–15\,cm were tested, corresponding to finite-distance regimes between $\lambda/2$ and $3\lambda$ at 60\,GHz. Four radii—4\,cm, 5\,cm, 8\,cm, and 15\,cm—were selected as representative cases, as they span the practical limits of workspace reachability, collision constraints, and measurement relevance.

\begin{table}[t!]
\centering
\caption{Experimental Scan Configurations and Coverage}
\begin{tabular}{cccccc}
\toprule
\rowcolor{gray!30}
\textbf{Scan Config.} & \textbf{Radius} & \textbf{$\theta_{\max}$} & \textbf{$\theta$ Count} & \textbf{Points} & \textbf{Coverage} \\
\midrule
4\,cm/$15^\circ$  & 4\,cm  & $0^\circ$--$60^\circ$ & 5 & 120 & 100\% \\
5\,cm/$10^\circ$  & 5\,cm  & $0^\circ$--$70^\circ$ & 7 & 252 & 100\% \\
8\,cm/$20^\circ$  & 8\,cm  & $0^\circ$--$60^\circ$ & 5 & 72  & 100\% \\
15\,cm/$15^\circ$ & 15\,cm & $0^\circ$--$60^\circ$ & 4 & 120 & 100\% \\
\bottomrule
\end{tabular}
\label{tab:IV}
\end{table}

The angular increments used for each configuration were selected as a practical
trade-off between measurement density and total scan duration. As summarized in
Table~\ref{tab:IV}, the 5\,cm scan employs fine $10^\circ$ sampling (seven polar samples
per azimuth), while the 8\,cm configuration uses a coarser $20^\circ$ increment (five
polar samples). The maximum polar angle for each radius reflects collision-aware
reachability: the 4\,cm scans are limited to $\theta \leq 60^\circ$, whereas the elevated DUT configuration at 5\,cm enables $\theta \leq 70^\circ$. The resulting number of spatial points ranges from 72 to 252, which directly determines total acquisition time given the fixed dwell duration at each pose. In Table~\ref{tab:IV}, ``$\theta$ Count'' denotes the number of discrete polar angles sampled between $0$ and $\theta_{\max}$ using an angular step size $\Delta\theta$. The total number of sampled points is given by $N = N_\phi N_\theta$, where $N_\phi = 360^\circ / \Delta\phi$ corresponds to the number of azimuthal samples and $N_\theta = \text{``$\theta$ Count''}$.


\begin{figure*}[!t]
    \centering
    \subfloat[]{\includegraphics[width=0.48\textwidth]{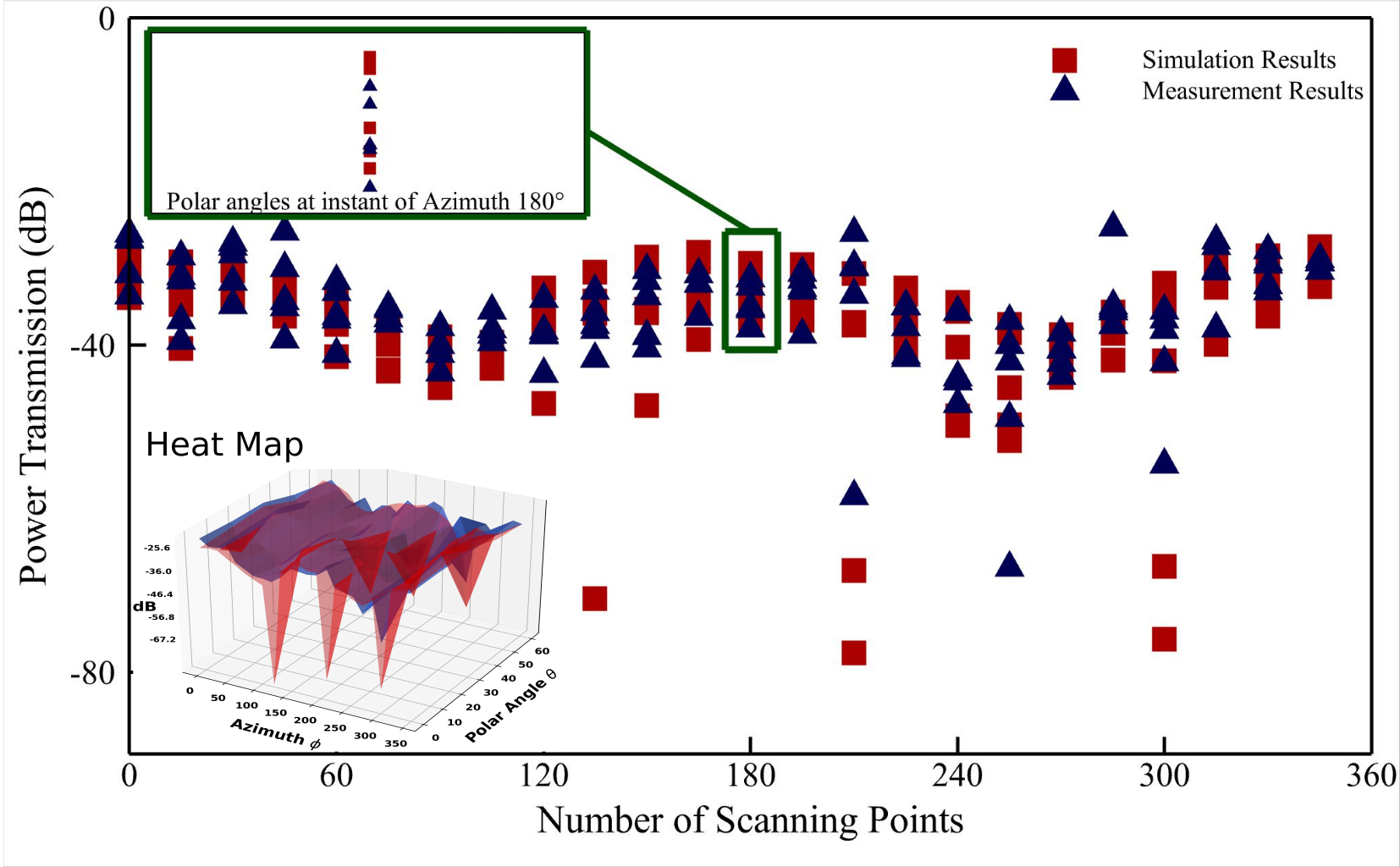}}
    \label{fig7a}
    \hfil
    \subfloat[]{\includegraphics[width=0.48\textwidth]{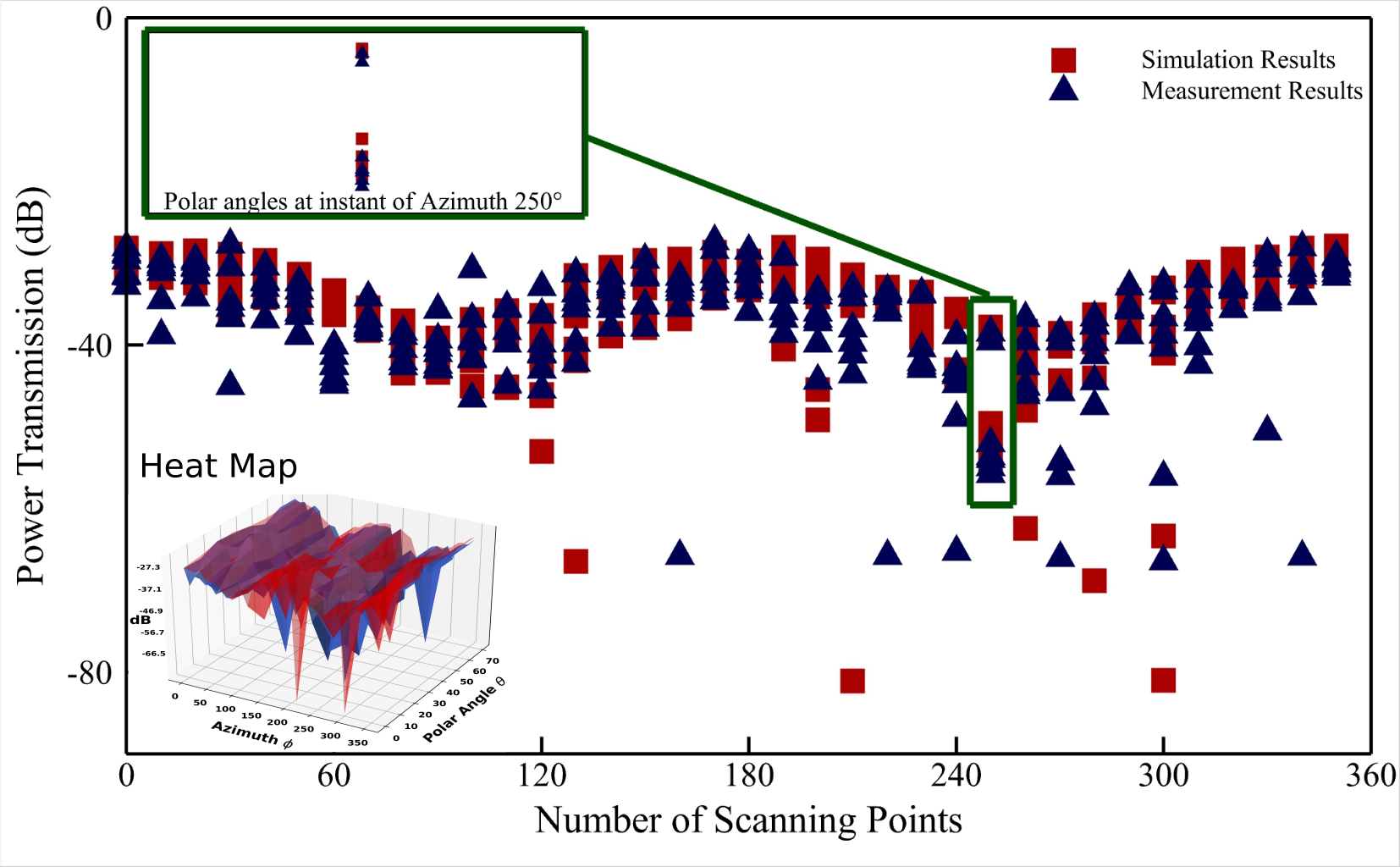}}
    \label{fig7b}

    \vspace{0.2cm}

    \subfloat[]{\includegraphics[width=0.48\textwidth]{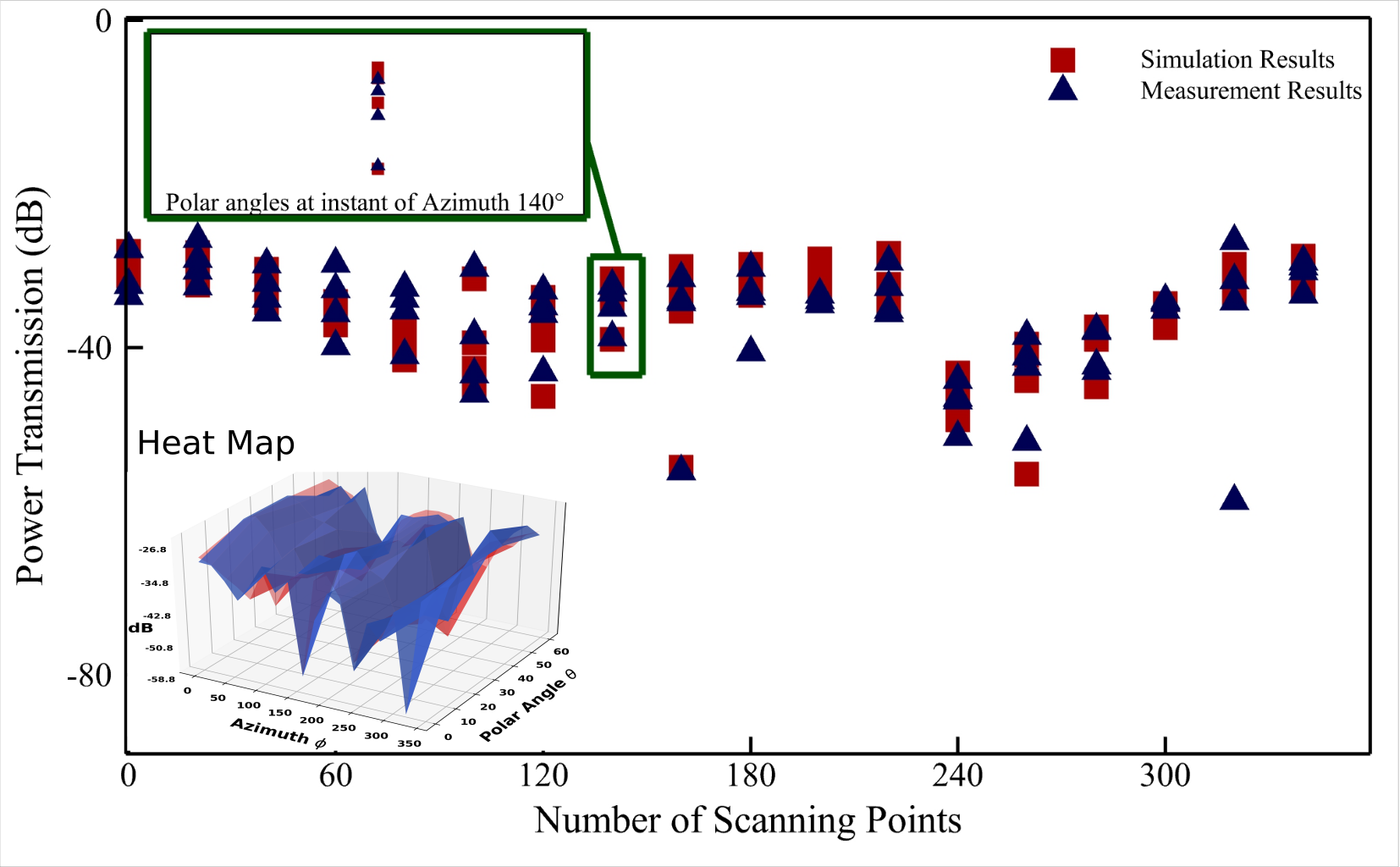}}
    \label{fig7c}
    \hfil
    \subfloat[]{\includegraphics[width=0.48\textwidth]{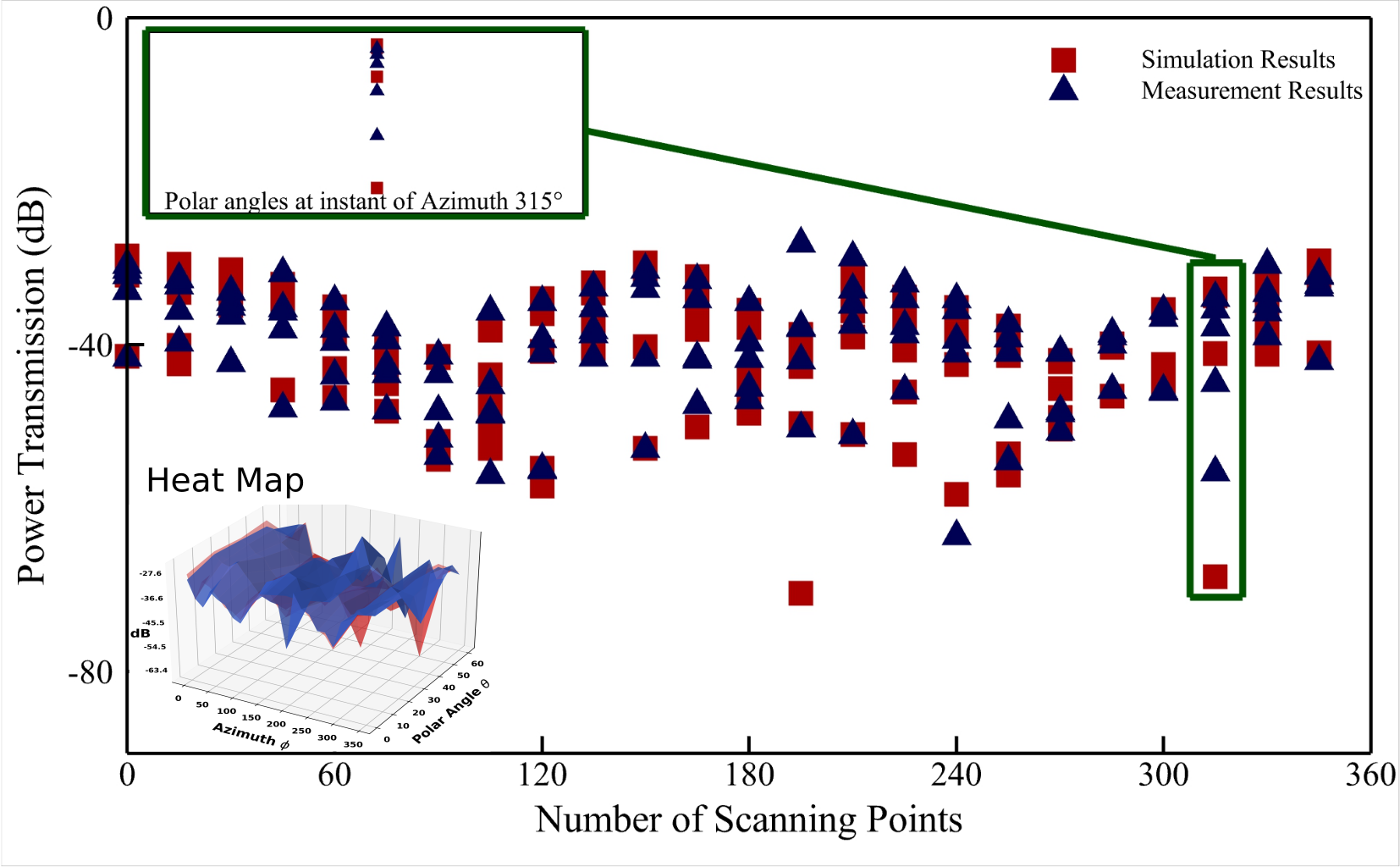}}
    \label{fig7d}

    \caption{Power-domain comparison between measured angular received power and the idealized free-space reference derived from full-wave simulation across the hemispherical sampling grid for four representative scan configurations:(a) $r = 4$\,cm with $\Delta\theta=\Delta\phi=15^\circ$, $\theta=0^\circ$--$60^\circ$, and full azimuthal sweep; (b) $r = 5$\,cm with $\Delta\theta=\Delta\phi=10^\circ$ and dense angular sampling up to $\theta=70^\circ$; (c) $r = 8$\,cm with $\Delta\theta=\Delta\phi=20^\circ$ and reduced sampling for mid-range coverage; and (d) $r = 15$\,cm with $\Delta\theta=\Delta\phi=15^\circ$, illustrating increasingly far-field--like angular received-power behavior at larger measurement radius.}
    \label{fig7}
\end{figure*}
\begin{table}[t!]
\centering
\small
\caption{Recurring Performance Metrics (Power-Domain)}
\label{tab:V}
\begin{tabularx}{\linewidth}{X X X X X X}
\toprule
\rowcolor{gray!30}
\textbf{Scan Config.} & \textbf{Repeat. (dB)} &
\textbf{Day-to-Day (dB)} & \textbf{Success (\%)} &
\textbf{s/pt} & \textbf{Scan Time (min)} \\
\midrule
4\,cm/$15^\circ$  & 0.18 & 0.24 & 100 & 20 & 40 \\
5\,cm/$10^\circ$  & 0.19 & 0.22 & 100 & 20 & 84 \\
8\,cm/$20^\circ$  & 0.17 & 0.19 & 100 & 20 & 24 \\
15\,cm/$15^\circ$ & 0.16 & 0.18 & 100 & 20 & 40 \\
\bottomrule
\end{tabularx}
\end{table}

Figures~\ref{fig7}(a)–(d) illustrate the executed hemispherical sampling grids for the
representative configurations summarized in Table~\ref{tab:IV}. Each subfigure shows
the discretized set of $(\phi,\theta)$ samples at the corresponding measurement radius
together with the achievable polar span $\theta_{\max}$ imposed by workspace and
collision constraints. The purpose of these figures is to visualize the planned
coverage and to verify that the system executes the commanded hemispherical sampling
set without violating workspace constraints. All configurations in
Table~\ref{tab:IV} achieved complete geometric coverage (100\%), confirming that the
motion-planning pipeline successfully reached every commanded pose across the tested
radii and angular resolutions. Coverage denotes the fraction of commanded sampling poses for which a valid motion plan and corresponding received-power measurement were successfully obtained.

\begin{figure}[t!]
    \centering
    \includegraphics[width=0.5\textwidth]{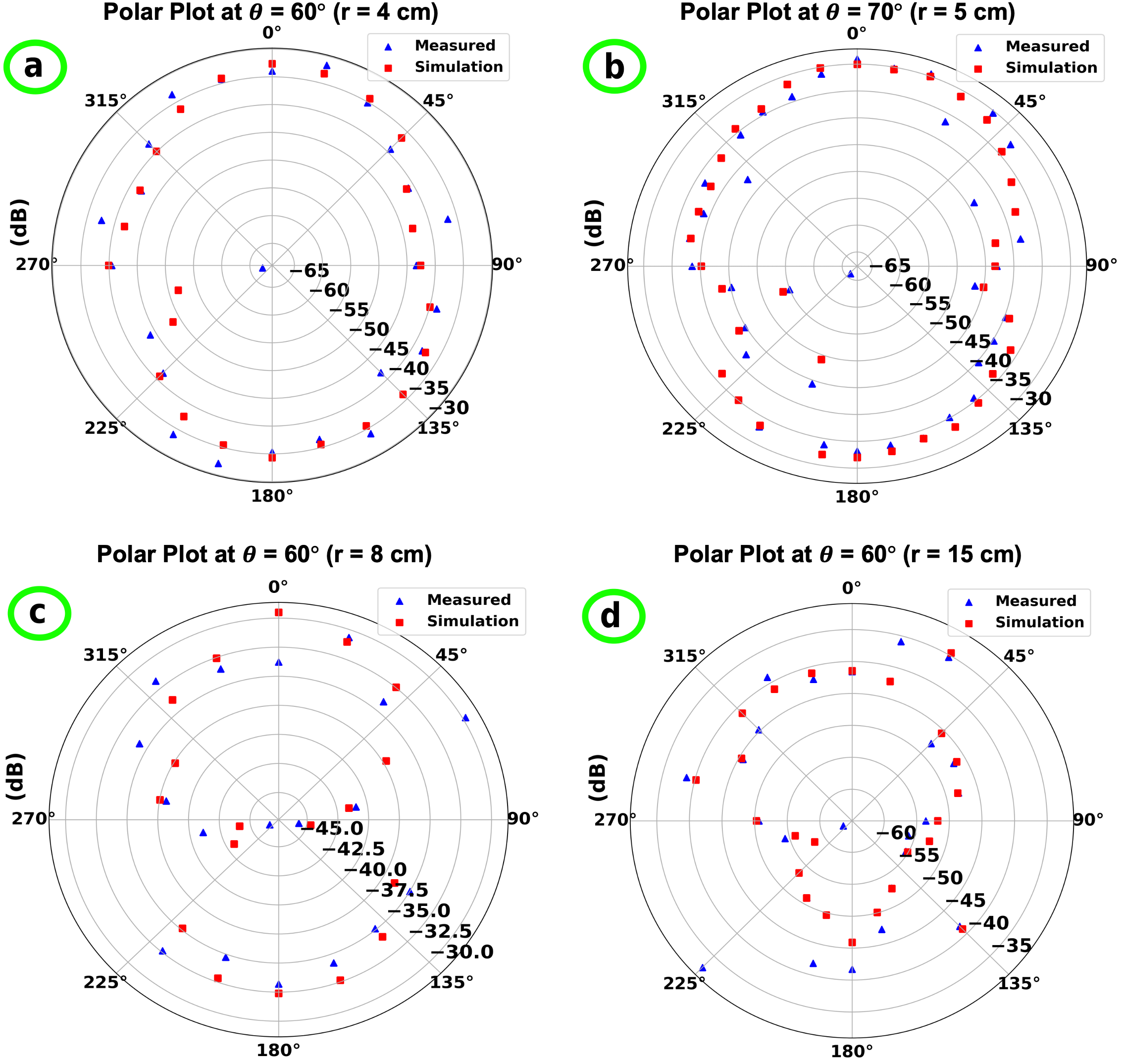}
    \caption{Representative power-domain angular received-power measurements fo edge case (maximum polar angle) obtained using the system, shown alongside idealized free-space reference responses derived from full-wave simulation for comparative evaluation.}
    \label{fig8}
\end{figure}

Complementing the geometric coverage analysis, Fig.~\ref{fig8} shows the corresponding power-domain angular characterization outcomes for the same representative cases. The four polar plots present measured azimuthal cuts of received power at fixed elevation angles and measurement radii, including finite-distance cases (e.g., $\theta=60^\circ$ at $r=4$~cm and $r=8$~cm) and an increasingly far-field--like configuration ($\theta=60^\circ$ at $r=15$~cm). Red markers denote the measured received-power values, and blue markers denote the simulation-derived power-domain reference evaluated at the same sampling directions. Across the evaluated radii, the measurements exhibit strong correlation with the reference in terms of peak direction and overall angular structure, with modest amplitude deviations. This agreement indicates that the geometry-calibrated hemispherical sampling and pose-indexed acquisition pipeline enables repeatable angular received-power characterization over multiple radii and supports consistent power-domain comparison against simulation references and manual baseline measurements.

\begin{table*}[t!]
\centering
\caption{Power-domain comparison against an idealized free-space simulation reference and a manual baseline.}
\begin{tabular}{ccccccccc}
\toprule
\rowcolor{gray!30}
\textbf{Scan Config.} & \textbf{Method} & \textbf{MAE (dB)} & \textbf{RMSE (dB)} &
\textbf{Std. Err. (dB)} & \textbf{SNR (dB)} & \textbf{$\rho$} &
\textbf{Max $P_{\mathrm{rx}}$ (dB)} & \textbf{Peak Dir. ($^\circ$)} \\
\midrule
4\,cm/$15^\circ$  & Automated & 1.41 & 1.71 & 1.17 & 39.5 & 0.988 & -26.3 & 15 \\
4\,cm/$15^\circ$  & Manual baseline & 2.22 & 2.72 & 1.31 & 33.1 & 0.945 & -26.6 & 15 \\
4\,cm/$15^\circ$  & Idealized free-space ref. & -- & -- & -- & -- & 1.000 & -26.1 & 15 \\
\midrule
5\,cm/$10^\circ$  & Automated & 1.58 & 1.87 & 1.21 & 37.2 & 0.981 & -27.8 & 10 \\
5\,cm/$10^\circ$  & Manual baseline & 1.67 & 2.05 & 1.72 & 31.6 & 0.938 & -28.0 & 10 \\
5\,cm/$10^\circ$  & Idealized free-space ref. & -- & -- & -- & -- & 1.000 & -28.1 & 10 \\
\midrule
8\,cm/$20^\circ$  & Automated & 1.23 & 1.43 & 0.98 & 41.7 & 0.995 & -29.0 & 20 \\
8\,cm/$20^\circ$  & Manual baseline & 1.79 & 1.94 & 1.50 & 35.2 & 0.964 & -29.4 & 20 \\
8\,cm/$20^\circ$  & Idealized free-space ref. & -- & -- & -- & -- & 1.000 & -29.6 & 20 \\
\midrule
15\,cm/$15^\circ$ & Automated & 1.19 & 1.34 & 0.91 & 44.0 & 0.997 & -30.0 & 15 \\
15\,cm/$15^\circ$ & Manual baseline & 1.52 & 2.37 & 1.20 & 36.4 & 0.971 & -30.2 & 15 \\
15\,cm/$15^\circ$ & Idealized free-space ref. & -- & -- & -- & -- & 1.000 & -30.4 & 15 \\
\bottomrule
\end{tabular}
\label{tab:VI}
\end{table*}

\subsection{Quantitative Accuracy, Robustness, and Repeatability}

To assess the power-domain measurement performance of the proposed methodology, each acquired angular received-power dataset was evaluated using two complementary reference points: (i) an idealized free-space reference derived from a full-wave finite-element electromagnetic solver, and (ii) conventional manual probe-station/turntable-style measurements representing a typical operator-dependent laboratory workflow. This dual-reference evaluation enables a structured assessment of how geometry-calibrated, automated spatial sampling performs relative to both an idealized simulation baseline and a practical manual baseline, without implying absolute metrological accuracy.

Beyond pointwise accuracy, a primary objective of the experimental campaign is to
evaluate operational robustness and repeatability under repeated hemispherical scanning.
To this end, multiple scans were conducted under identical geometric configurations
within a single session (intra-day), as well as across different days following full
system reinitialization. These experiments bound the sensitivity of the received-power
measurements to calibration drift, mechanical variability, environmental conditions,
and long-duration operation in a realistic laboratory setting. The evaluation protocol computes robustness and operational metrics from received-power measurements acquired over the full $(\phi,\theta)$ sampling grid and aggregates them across three independent trials per configuration to capture statistical stability. Table~\ref{tab:V} summarizes these robustness metrics for four representative scan radii. The reported seconds-per-point correspond to the enforced dwell time $t_{\mathrm{dwell}}$ at each pose and do not include motion or communication overhead.

Here, repeatability denotes the mean absolute difference in received power across
repeated scans acquired under identical scan configurations within a single session
(intra-day), while the day-to-day metric quantifies the same measure across different
days following full system re-initialization. Additional metrics including planning
success rate, per-point acquisition time, total scan duration, and sampling
throughput characterize the overall operational efficiency of the automated measurement
workflow. Reported scan times correspond to dwell-only acquisition time based on the
specified seconds-per-point and exclude planning and setup overhead.

Having established robustness and operational consistency, we next evaluate the power-domain accuracy of the proposed methodology relative to reference measurements. For each scan configuration, standard deviation–based accuracy metrics were computed over the sampled $(\phi,\theta)$ grid, including mean absolute error (MAE), root-mean-square error (RMSE), standard error, and peak received power. These metrics capture complementary aspects of deviation, including average mismatch, sensitivity to outliers, and distribution spread. In addition, angular-shape agreement was quantified using the Pearson correlation coefficient $\rho$ between the measured and reference received-power vectors evaluated over the sampling grid, while the peak-response direction was extracted to assess angular consistency. All reported metrics are evaluated exclusively in the received-power domain. Table~\ref{tab:VI} reports the resulting power-domain accuracy and agreement metrics for the automated measurement system, the manual baseline, and the idealized free-space reference. This comparison enables direct assessment of angular trend correlation, amplitude deviation, and peak-direction consistency across the different measurement modalities.

To complement aggregate accuracy metrics such as MAE and RMSE, an error-distribution analysis was conducted and reported in Table~\ref{tab:VI} to characterize how received-power deviations vary across the hemispherical sampling domain. Specifically, pointwise deviations are evaluated as $\Delta(\phi,\theta)=P_{rx}^{\mathrm{meas}}(\phi,\theta,r)-P_{rx}^{\mathrm{ref}}(\phi,\theta,r)$ in the power domain (dB) over the sampled $(\phi,\theta)$ grid at a fixed radius $r$, and their statistical distributions are summarized using MAE, RMSE, standard error, and correlation metrics. This distributional perspective highlights angular regions in which agreement is more sensitive to finite-distance effects, reduced signal-to-noise ratio, or workspace boundary constraints, particularly near the maximum accessible polar angle $\theta_{\max}$. In addition to quantifying average accuracy, Table~\ref{tab:VI} therefore provides insight into the spatial consistency of the measurements and the angular dependence of residual discrepancies.

For each scan configuration, pointwise deviations were computed over the full set of
$(\phi,\theta)$ sampling locations and summarized using the median, standard deviation,
and extreme values (minimum and maximum). These statistics, reported in
Table~\ref{tab:VII}, provide a compact description of the spread and concentration of
received-power deviations without assuming a specific underlying distribution.
Including extreme values helps identify edge-case behavior, such as samples acquired
near workspace limits or in regions where reduced mechanical clearance increases
sensitivity to small positioning variations.

\begin{table}[t!]
\centering
\setlength{\tabcolsep}{5pt}
\caption{Distribution of absolute received-power deviations (in dB) relative to the idealized free-space simulation reference across the sampled $(\phi,\theta)$ grid.}
\begin{tabular}{ccccc}
\toprule
\rowcolor{gray!30}
\textbf{Scan Config.} & \textbf{Min (dB)} & \textbf{Median (dB)} & \textbf{Std. Dev. (dB)} & \textbf{Max (dB)} \\
\midrule
4\,cm/$15^\circ$  & 0.03 & 1.66 & 1.54 & 5.2 \\
5\,cm/$10^\circ$  & 0.04 & 1.85 & 1.78 & 6.1 \\
8\,cm/$20^\circ$  & 0.02 & 1.18 & 1.07 & 3.5 \\
15\,cm/$15^\circ$ & 0.01 & 1.11 & 1.02 & 2.9 \\
\bottomrule
\end{tabular}
\label{tab:VII}
\end{table}

Figure~\ref{fig8} further illustrates representative edge-case conditions for each scan configuration by visualizing selected angular cuts of the measured received-power response. These plots highlight angular regions in which deviations are most sensitive to geometric factors such as finite sampling resolution, reduced clearance, or proximity to the limits of the accessible workspace. Together with the tabulated statistics, these visualizations provide a complementary perspective for interpreting how received-power deviations are distributed across the hemispherical sampling domain under realistic measurement conditions.

\vspace{2mm}

\section{Results and Discussion}
\label{sec:VI}

The full-wave simulation results are used in this work as an idealized free-space reference rather than as a representation of the experimental environment. The finite-element model intentionally excludes surrounding structures, cabling, robotic components, and laboratory boundaries in order to provide a controlled baseline. Accordingly, discrepancies between measurement and simulation are expected due to environmental scattering, absorber limitations, interactions with the robotic structure, and cabling effects that are not included in the simulation model. The objective of the comparison is therefore not absolute agreement, but assessment of angular-trend correlation and repeatability under realistic measurement conditions. All synchronization in this work is temporal and procedural, ensuring consistent pose-to-measurement association rather than electromagnetic phase alignment.

The experimental campaign spans thirteen hemispherical scan configurations and more than twenty full scan sessions, enabling a comprehensive assessment of the proposed methodology in terms of geometry-calibrated spatial fidelity, power-domain agreement with an idealized reference, and operational robustness for unattended multi-hour acquisition. The key outcome is that repeatable, angle-indexed mmWave received-power characterization can be achieved without turntables or dedicated anechoic fixtures by combining deterministic pose generation, collision-aware execution, and synchronized received-power readout under realistic installation constraints.

Across all tested radii (3--15\,cm) and angular resolutions (10$^\circ$--20$^\circ$), the system achieved 100\% spatial coverage of the commanded hemispherical grids with no collisions, emergency stops, or aborted trajectories. This reliability is notable because the most demanding cases occur at small radii and high polar angles, where clearance is minimal and joint-limit proximity can introduce local infeasibility. The planned grids in Fig.~\ref{fig7}(a)--(d) confirm that dense hemispherical sampling remains reachable even in constrained near-range configurations, supporting the practicality of the geometry-calibrated spatial sampling strategy over a wide range of installation-dependent test setups.

The motion-planning benchmark in Table~\ref{tab:3} further clarifies why dense hemispherical sampling remains feasible at scale. RRT-Connect provides the most favorable trade-off for repeated trajectory generation, achieving an average planning time of 0.45\,s with 100\% success and competitive path lengths, which is advantageous when planning is invoked repeatedly over large scan grids. BIT* and CHOMP offer modest improvements in path properties but require substantially longer computation times (1.5--3.2\,s), while the PPO planner demonstrates intermediate performance that suggests potential for future adaptation in dynamically changing environments. Importantly, any observed reductions in measurement error relative to the manual baseline are attributed primarily to reduced alignment variability and consistent spatial sampling; these improvements reflect procedural consistency rather than absolute metrological superiority over calibrated chamber.

Robustness is quantified directly through repeated scans under identical configurations. As summarized in Table~\ref{tab:V}, intra-day repeatability remains below 0.20\,dB and day-to-day variation remains below 0.25\,dB across the representative scan cases. These results indicate that the calibrated transformation chain, end-effector mounting repeatability, and receiver chain remain stable across repeated acquisitions and power-cycle resets, which is essential for producing meaningful power-domain comparisons across radii and across independent measurement campaigns.

With this operational stability established, Table~\ref{tab:VI} reports power-domain comparison metrics against the idealized free-space reference and the manual baseline workflow. Across all evaluated radii, the automated system achieves MAE values between 1.19\,dB and 1.58\,dB, with RMSE below 1.9\,dB. The Pearson correlation coefficient remains above 0.98 for all cases, indicating strong global agreement in angular received-power trends, while the manual baseline exhibits higher mean errors and reduced correlation, consistent with operator-dependent variability in probe
placement and orientation. The resulting reduction in error (up to 36.5\%) is therefore explained primarily by deterministic spatial sampling and reduced alignment variability, which tighten the pose-to-measurement consistency across the entire hemispherical grid. In addition, the extracted peak-response directions match the simulation reference within the imposed sampling resolution, indicating that the geometry-calibrated transformation chain maintains stable boresight alignment throughout multi-axis trajectories.

Beyond aggregate metrics, the distributional statistics in Table~\ref{tab:VII} provide insight into how deviations are distributed across the hemisphere. Median deviations remain below 2\,dB for all representative configurations, while maximum deviations range from 2.9\,dB to 6.1\,dB. Higher deviations appear primarily in expected regimes: near the accessible boundary ($\theta \rightarrow \theta_{\max}$), where the manipulator operates closer to joint limits and is more sensitive to small orientation perturbations, and in low-received-power angular sectors where reduced SNR increases sensitivity to receiver noise and environmental scattering. Figure~\ref{fig8} illustrates representative angular cuts in which these edge-case sensitivities appear. Importantly, the results do not exhibit abrupt discontinuities or isolated outliers, supporting the stability of the pose-indexed acquisition pipeline over the full scan domain.

From a practical deployment standpoint, throughput is dominated by the RF dwell and trace stabilization requirements rather than robot motion. With a per-point acquisition time of $\approx 20$\,s, total scan time scales approximately linearly with the number of sampled points, yielding scan durations of 24--84\,min for the representative configurations in Table~\ref{tab:V}. No data loss requiring manual intervention occurred during the reported campaigns. The fallback-to-home replanning mechanism was invoked only under boundary conditions and resolved feasibility
deterministically without materially impacting scan completion.

Overall, these results establish that the proposed methodology provides: (i) reliable execution of commanded hemispherical sampling (100\% coverage), (ii) strong repeatability under repeated scans ($<0.20$\,dB intra-day and $<0.25$\,dB day-to-day), (iii) power-domain agreement with an idealized free-space reference (MAE $<2$\,dB and $\rho>0.98$), and (iv) consistent angular received-power characterization across finite measurement radii exhibiting increasingly far-field--like behavior as the radius increases. Together, these outcomes support the use of the proposed system as a practical and portable alternative to operator-dependent manual workflows for engineering validation and comparative characterization of embedded mmWave transmitters under realistic installation constraints.


\section{Conclusions}
\label{sec:VII}

This work presented an automated measurement methodology for angular received-power characterization of embedded mmWave transmitters based on geometry-calibrated spatial sampling and pose-indexed RF acquisition. Collaborative robot motion planning is employed solely as an enabling tool to achieve controlled, repeatable probe positioning under realistic installation constraints, while collision-aware planning and quasi-static pose stabilization support unattended hemispherical sampling in a semi-anechoic, absorber-augmented environment. Across all evaluated radii and angular resolutions, the system achieved complete coverage of the commanded sampling grids with high repeatability and operational robustness. Power-domain validation against idealized free-space references derived from full-wave simulation, together with comparisons against operator-dependent manual baseline measurements, demonstrates improved alignment consistency and repeatability under realistic laboratory conditions. Reductions of up to 36.5 \% in mean absolute error relative to manual workflows are attributed primarily to deterministic spatial sampling and reduced pose variability rather than to changes in electromagnetic measurement modality. Importantly, the proposed framework is not intended to replace classical antenna metrology, nor to perform probe-corrected or phase-resolved field reconstruction. Instead, it addresses a complementary and increasingly common validation need: repeatable, angle-indexed received-power characterization of embedded active mmWave modules in situ, where isolation, turntables, and fully anechoic facilities are unavailable.
Because the method relies only on geometry-calibrated spatial sampling and amplitude-only received-power acquisition, it is readily extensible to a wide range of embedded radar platforms without loss of generality, including automotive radars integrated behind vehicle bumpers, radar-equipped drones, robotic platforms, and wearable or body-mounted sensing systems. The demonstrated repeatability, scalability, and deployment flexibility establish the proposed methodology as a practical tool for engineering validation and comparative characterization during system integration. Future work will focus on extending the framework to coherent field measurements, adaptive scan strategies, and multi-robot configurations for larger or more complex installed platforms.

\section{Acknowledgments} \label{sec:VIII}
The authors gratefully acknowledge the support of Aidin Taeb and Keysight Technologies for providing the millimeter-wave receiver (Rx), harmonic mixer, and signal analyzer equipment used in this study. Their technical guidance and generous equipment support were instrumental to the successful development and validation of the system. This work was also supported by Infineon, Google, Rogers, and MITACS.


\bibliographystyle{IEEEtran}
\bibliography{Ref}

\begin{IEEEbiography}[{\includegraphics[width=1.1in,height=1.3in,clip,keepaspectratio]{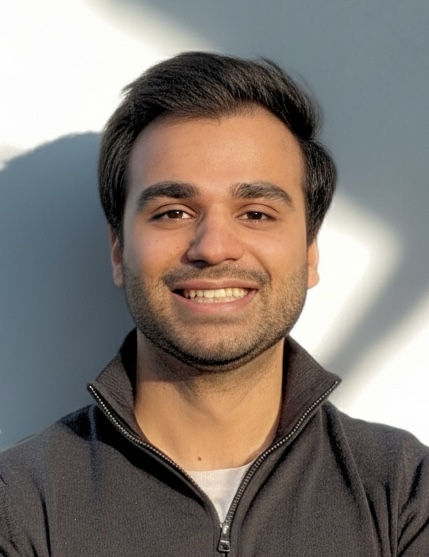}}]{Maaz Qureshi}
(Graduate Student Member, IEEE) received the M.A.Sc. degree in Mechanical and Mechatronics Engineering from the University of Waterloo, ON, Canada. He is currently working in industry as an AI and Robotics Software Engineer, where he develops autonomous robotic systems and perception-driven AI/ML algorithms for real-world industrial applications.

His research spans applied machine-learning inference and framework development, multi-robot autonomy, distributed simultaneous localization and mapping (SLAM), 5G-connected robotic systems, 4D mmWave radar perception, and collaborative robotic manipulation. He has contributed to publications in IEEE robotics conferences and journals, with work encompassing computational intelligence, visual perception, and swarm robotics. He has also contributed professionally in the underwater robotics industry, where remotely operated and autonomous underwater vehicles (ROVs/AUVs) perform autonomous structural inspection and mapping using sonar and high-fidelity stereo camera–based sensor fusion for imaging and crack propagation analysis.
\end{IEEEbiography}

\begin{IEEEbiography}[{\includegraphics[width=1.1in,height=1.3in,clip,keepaspectratio]{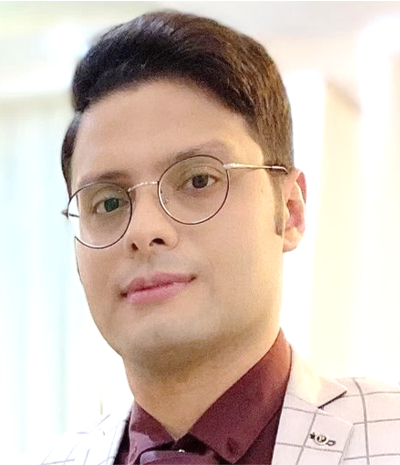}}]{Mohammad Omid Bagheri} (Member, IEEE) received the B.Sc. degree in Electrical Engineering from the Amirkabir University of Technology, Tafresh Branch, Iran, in 2012, the M.Sc. degree in Telecommunications Engineering from Shahed University, Tehran, Iran, in 2016, and the Ph.D. degrees in Telecommunications Engineering from Shahed University, Tehran, Iran, in 2021, and in Antennas, Microwaves, and Wave Optics from the University of Waterloo, ON, Canada, in 2025. He is currently a Postdoctoral Fellow with the Department of Electrical and Computer Engineering, University of Waterloo, and is also affiliated with the Wireless Sensors and Devices Laboratory. 

Dr. Bagheri’s research has resulted in high-impact publications in Nature Communications, IEEE journals and conferences, with a focus on metasurface-enhanced antennas, radar sensing technologies, and biomedical RF applications. These contributions have been recognized through several prestigious honors, including the Ph.D. Student Initiative Award at the IEEE IMBioC 2023 in Belgium, the IEEE APS Honorable Mention Student Paper Award at AP-S/URSI 2024 in Florence, Italy, Best Student Paper Award at NEMO 2024 in Montreal, Canada, a Paper student contest Finalist at iWAT 2025, and leading the finalist team at the IEEE AP-S Student Design Contest 2025. His work on non-invasive blood glucose monitoring for next-generation smartwatches was featured on CTV News Canada, and his innovation in radar-based biosensing earned a Gold Medal and multiple special awards at the iCAN 2024 Competition. His article was ranked 7th among the 170 most accessed papers in Nature Communications in 2024. His contributions have been recognized with a U.S. patent and the Entrepreneurial PhD Fellowship at the Conrad School of Entrepreneurship and Business, supporting his efforts to merge engineering expertise with business strategy to drive tech innovation.
\end{IEEEbiography}

\begin{IEEEbiography}[{\includegraphics[width=1.1in,height=1.3in,clip,keepaspectratio]{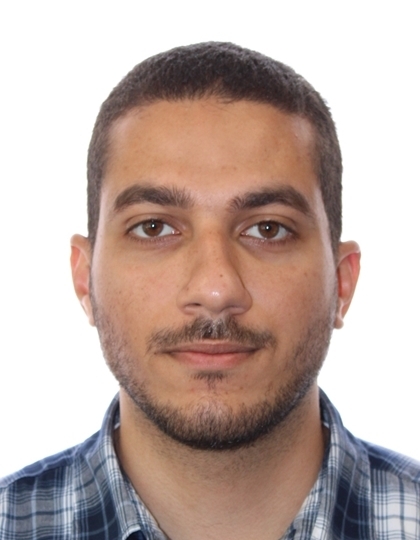}}]{Abdelrahman Elbadrawy} received his MASc degree in Antennas, Microwaves, and Wave Optics from the Department of Electrical and Computer Engineering at the University of Waterloo, ON, Canada, in 2024. He was a member of the Wireless Sensors and Devices Laboratory. He received his B.Sc. degree in Communications and Information Engineering from the University of Science and Technology at Zewail City in 2022.
His research focuses on the development of digital twins for radar and joint communication–sensing systems.
\end{IEEEbiography}

\begin{IEEEbiography}[{\includegraphics[width=1in,height=1.25in,clip,keepaspectratio]{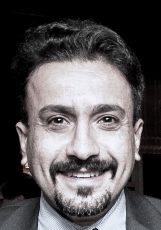}}]{William Melek}  (M’02–SM’06) received the
M.A.Sc. and Ph.D. degrees in mechanical engineering from the University of Toronto, Toronto, ON, Canada, in 1998 and 2002, respectively.
Between 2002 and 2004, he was the Artificial Intelligence Division Manager with Alpha Global IT, Inc., Toronto. He is currently a Director of UW ROBOHUB lab and Professor with the Department of Mechanical and Mechatronics Engineering, University of Waterloo, ON, Canada. His current research interests include Operational Artificial Intelligence, Intelligent Control, Computational Intelligence and Advanced robotics, neural networks, and genetic algorithms for modeling and control of dynamic systems.
Dr. Melek is a member of the American Society of Mechanical Engineers.
\end{IEEEbiography}

\begin{IEEEbiography}[{\includegraphics[width=1in,height=1.25in,clip,keepaspectratio]{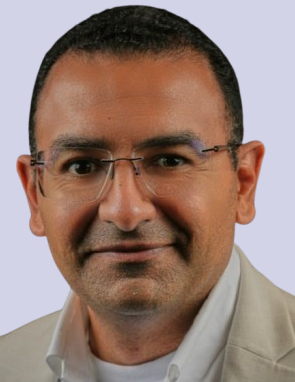}}]{George Shaker} (S IEEE 1997, SM IEEE 2015) is the lab director of the Wireless Sensors and Devices Laboratory at the University of Waterloo, where he is an adjunct associate professor at the Department of Electrical and Computer Engineering. Previously, he was an NSERC scholar at Georgia Institute of Technology. Dr. Shaker also held multiple roles with RIM’s (BlackBerry). He is also the Chief Scientist at Spark Tech Labs, which was co-founded in 2011. With over twenty years of industrial experience in technology, and more than ten years as a faculty member leading project related to the application of wireless sensor systems for healthcare, automotives, and unmanned aerial vehicles, Prof. Shaker has many design contributions in commercial products available from startups and multinationals. A sample list includes Google, COM DEV, Honeywell, Blackberry, Spark Tech Labs, Bionym, Lyngsoe Systems, ON Semiconductors, Ecobee, Medella Health, NERV Technologies, Novela, Thalmic Labs, North, General Dynamics Land Systems, General Motors, Toyota, Maple Lodge Farms, Rogers Communications, and Purolator.

Dr. Shaker has authored/coauthored 200+ publications and 35+ patents/patent applications. George has received multiple recognitions and awards, including the the IEEE AP-S Best Paper Award (the IEEE AP-S Honorable Mention Best Paper Award (4 times to-date), the IEEE Antennas and Propagation Graduate Research Award, the IEEE MTT-S Graduate Fellowship, the Electronic Components and Technology Best of Session paper award, and the IEEE Sensors most popular paper award. Four papers he co-authored in IEEE journals were among the top 25 downloaded papers on IEEEXplore for several consecutive months. He was the supervisor of the student team winning the third best design contest at IEEE AP-S 2016 and 2025, co-author of the ACM MobileHCI 2017 best workshop paper award, and the 2018 Computer Vision Conference Imaging Best Paper Award. He co-received with his students several research recognitions including the NSERC Top Science Research Award 2019, IEEE APS HM paper award (2019, 2022, 2023, 2024, and 2025), Biotec top demo award 2019, arXiv top downloaded paper (medical device category) 2019, Velocity fund 2020, NASA Tech Briefs HM Award (medical device category) 2020, UW Concept 2021, UK Dragons Canadian Competition 2021, CMC Nano 2021, COIL COLAB 2022, Wiley Engineering Reports top downloaded paper for 2022, Canadian Space Agency Cubesat Design winner 2023, IEEE NEMO Best Paper Award 2024, Nature Communications Engineering Top 25 downloaded papers in 2024, IEEE MAPCON Best Paper Award 2024, iWAT 2025 paper finalist, and Prototypes for Humanity 2025 finalist.

Dr. Shaker is currently the Chair of the IEEE AP-S Membership \& Benefits Committee, Chair of the IEEE AP-S TC-10 on Environment, Sustainability, and Societal Impacts, Vice-Chair of the IEEE MTT-S TC-27 on Autonomous Systems, and an IEEE Sensors Council Distinguished Lecturer.
\end{IEEEbiography}

\end{document}